\RequirePackage[british]{babel}
\pdfoutput=1
\documentclass[reqno,a4paper,12pt]{amsart}
\usepackage[a4paper,hmargin=2cm,tmargin=3cm,bmargin=3cm]{geometry}
\usepackage[utf8x]{inputenc}
\usepackage[T1]{fontenc}
\usepackage{mathptmx}
\usepackage{amsmath,amssymb,amstext,amsthm,amscd,mathrsfs,eucal,bm,slashed}
\usepackage{graphicx,color}
\usepackage{array}
\usepackage{cite}
\usepackage{hyperref}
\hypersetup{%
  pdftitle   = {Bondi mass},
  pdfkeywords = {Bondi mass, asymptotic flatness, higher dimensions, Killing spinors},
  pdfauthor  = {S. Hollands, A. Thorne},
  pdfcreator = {\LaTeX\ with package \flqq hyperref\frqq}
}

\newcommand{\half}{\tfrac12}

\newcommand{\RR}{\mathbb{R}}
\newcommand{\CC}{\mathbb{C}}

\newcommand{\eH}{\mathscr{H}}

\newcommand{\cK}{\mathscr{K}}

\newcommand{\cS}{\mathscr{S}}

\theoremstyle{plain}
\newtheorem{lemma}{Lemma}

\newtheorem{theorem}[lemma]{Theorem}

\theoremstyle{definition}

\newtheorem{remark}[lemma]{Remark}

\newcommand{\MUNCH}[1]{\relax}

\allowdisplaybreaks
\def\ben{\begin{equation}}
\def\een{\end{equation}}
\def\bena{\begin{eqnarray}}
\def\eena{\end{eqnarray}}
\def\non{\nonumber}

\renewcommand{\half}{\tfrac{1}{2}}

\newcommand{\D}{{\mathcal D}}

\newcommand{\M}{\mathscr{M}}

\renewcommand{\d}{\mbox{d}}

\newcommand{\mr}{\mathbb{R}}

\newcommand{\I}{{\mathscr{I}}}
\begin{document}
\title[Bondi mass]{Bondi mass cannot become negative in higher dimensions}
\author[Hollands]{Stefan Hollands}
\author[Thorne]{Alexander Thorne}
\address{Universit\"{a}t Leipzig, Institut f\"{u}r Theoretische Physik, Br\"{u}derstrasse 16, D-04103 Leipzig, FRG}
\email{HollandsS@cardiff.ac.uk}
\address{School of Mathematics, Cardiff University, Senghennydd Road, Cardiff CF24 4AG, Wales, UK}
\email{ThorneA@cardiff.ac.uk}
\date{\today}
\vspace*{-.7cm}
\begin{abstract}
We prove that the Bondi mass of an asymptotically flat, vacuum, spacetime
 cannot become negative in any even dimension $d \ge 4$. The notion of Bondi mass is more subtle in $d > 4$ dimensions because radiating metrics have a slower decay than stationary ones, and those subtleties are reflected by a considerably more difficult proof of positivity. Our proof holds for the standard spherical infinities, but also extends to infinities of more general type which are $(d-2)$-dimensional manifolds admitting a real Killing spinor. Such manifolds typically have special holonomy and Sasakian structures. The main technical advance of the paper is
an expansion technique based on ``conformal Gaussian null coordinates''. This expansion helps us to understand the
consequences imposed by Einstein's equations
on the asymptotic tail of the
metric as well as auxiliary spinorial fields.
As a by-product, we derive a coordinate expression for the geometrically invariant formula for the Bondi
mass originally given by Hollands and Ishibashi.
\end{abstract}
\maketitle

\section{Introduction}
\label{sec:introduction}

In general relativity, there are two notions of mass of an asymtptocially flat spacetime.
The ADM mass is defined at spatial infinity. It measures the total mass of an initial
data set on an asymptotically flat slice, i.e. a slice of constant $t$ in an asymptotically Cartesian coordinate system.
The Bondi mass is defined at null-infinity. It measures the total mass associated with
an asymptotically hyperboloidal slice, i.e. a slice approaching constant retarded time $u=t-R,\ R=(\sum x_i^2)^{1/2}$ in an asymptotically Cartesian coordinate system. The ADM mass is independent of $t$, but the Bondi mass is in general a function of $u$, or more precisely, of a cut of null-infinity. Its change reflects an outgoing flux of gravitational radiation.

Both notions of mass were first defined for 4-dimensional spacetimes. While the ADM-mass is
readily generalized to arbitrary spacetime dimension $d \ge 4$, a generalization of the Bondi mass required considerably more work~\cite{hi}, see also~\cite{Tanabe:2011es,tan} for
a different approach. The root cause of the difficulty can roughly be seen from the following consideration. Near spatial infinity, the deviation of a non-trivial vacuum solution from Minkowski space is of order $R^{-d+3}$ for $R \to \infty$ as $t$ is held fixed. This behavior is seen explicitly e.g. for the Schwarzschild metric and corresponds to the decay of the potential of a point mass in Newtonian gravity. By contrast, near null-infinity, the deviation from Minkowski space is typically of order $R^{-d/2+1}$ for $R \to \infty$ as $u$ is held fixed. This behavior can  be seen crudely from the graviton propagator in Minkowski space. Thus, $d=4$ is special because both decays are of the same order, but for $d>4$,
the decay at null-infinity is slower. Nevertheless, the notion of Bondi mass corresponds to
terms deviating from the Minkowski metric of order $R^{-d+3}$ in any dimension, i.e. terms that are sub-leading, and are thus, in a sense, ``buried deep within the asymptotic expansion'' of the metric for large $d$. For this reason, both the definition of Bondi mass, as well as the investigation of its properties is considerably more subtle in dimensions
$d>4$. 

The first purpose of this paper is to investigate in much more detail the asymptotic expansion of the metric 
in $d\ge 4$ dimensions. A payoff of these investigations is going to be that  the notion of  Bondi-mass can
be defined under much less stringent assumptions than originally imposed in~\cite{hi}. We then combine our 
asymptotic expansion techniques with spinor methods and show that, under natural assumptions, the 
Bondi-mass is always non-negative. This result generalizes classical results in $d=4$ obtained previously 
by~\cite{Schon:1982re},~\cite{Horowitz:1981uw},~\cite{Ludvigsen:1981gf}, see also~\cite{Chrusciel:2003gn}.

Our plan for this paper is as follows. In the next section, we describe our technical assumptions, and state our results. 
Section~\ref{sec3} contains the proofs, but some lengthy formulas have been moved to appendix~\ref{appA}.  

\section{Assumptions and results}
\label{sec2}

Let us begin by stating carefully our basic assumptions regarding the asymptotics of the spacetime metric at null-infinity. 
We basically use the framework of conformal infinity \` a la Penrose~(see e.g.~\cite{wald}). This requires that $d$ is {\em even}~\cite{Hollands:2004ac}, which
is from now on tacitly assumed. In that framework, a spacetime $(\M,g)$ is asymptotically flat\footnote{We use the
term in a broader sense than usual. In particular, our assumptions do not necessarily imply that the metric
approaches a flat metric near infinity, or even that the topology of $\M$ is
that of $\mr^{d}$ outside a large compact set.} near
null infinity, if (i) there exists a conformal embedding $\M \to \tilde \M$ into a manifold
$\tilde \M$ with boundary $\I = \I^+ \cup \I^-$, and $\I^\pm \cong \RR \times \Sigma$ for compact $(d-2)$-dimensional manifold $\Sigma$. 
(ii) Under the conformal embedding, $g = \Omega^{-2} \tilde g$, where $\Omega$ is smooth up to and including the boundary and 
$\Omega = 0, \d \Omega \neq 0$ on $\I$. We also require (iii) that the metric $g$ is Ricci-flat, although our arguments would be practically unaffected by the presence
of a stress tensor with sufficient decay (which can easily be worked out from our proofs),
satisfying the energy conditions needed for positivity of mass. Using standard arguments this implies that $\I$ is a null surface. 

There is evidently a great deal of arbitrariness in the choice of $\Omega$, and hence in that of the unphysical metric $\tilde g$. A key idea of this paper is to (partially) remove this ambiguity in the following way. Near $\I^+$, we choose Gaussian null coordinates~\cite{Penrose} for the unphysical metric $\tilde g$ based on a
cut $\Sigma$ of $\I^+$. One can show~\cite{thorne} that the conformal factor $\Omega$ can be chosen such that it coincides with the Gaussian null-coordinate `$r$', so that, in an open neighborhood of
$\I^+$ we can write the metric as\footnote{The argument in~\cite{thorne} only establishes
that form with $r^2 \alpha$ replaced by $rf$. That $f$ vanishes on $\I^+$ is seen from the Einstein equation~\eqref{eq:ei-uu}.}
\ben\label{gform}
g = r^{-2} \left( 2 \, \d u(\d r - r^2\alpha \, \d u - r\beta_A \, \d x^A)
+ \gamma_{AB} \, \d x^A \d x^B \right) = r^{-2} \, \tilde g \ .
\een
We refer to this coordinate system as ``conformal Gaussian null coordinates'', or CGNC~\cite{Ishibashi:2007kb}\footnote{
This type of coordinate system has also been considered previously in $d=4$ in~\cite{tafel}.
}.
They are uniquely fixed by the choice of the cut $\Sigma$, up to a rescaling of $u,r$,
and obviously up to the unimportant choice of local coordinates $x^A,\ A=1, \dots, d-2$ on $\Sigma$. Conformal null infinity $\I^+$ is located at $r=0$, and $u$ is a coordinate along $\I^+$. The cut $\Sigma$ corresponds to
$r=0=u$. The usefulness of CGNC can largely be traced back to their geometrical origin:
$\partial/\partial r$ is tangent to an affinely parameterized congruence of null geodesics
transversal to $\I^+$, $\partial/\partial u$ is tangent to affinely parameterized null geodesics
ruling $\I^+$. The form~\eqref{gform} of the metric near $\I^+$ is {\em equivalent} to requirements (ii), and we will from now on work with this form. In $d$-dimensional Minkowski space, the infinity $\Sigma \cong S^{d-2}$ is spherical, and the change of coordinates $r=R^{-1},u=t-R$, with $x^A$ polar angles
on $S^{d-2}$ (endowed with the round metric $\gamma_{AB}$) brings the metric into the form~\eqref{gform}. Thus, $u$ is interpreted as a retarded time and $r$ is the
inverse distance at fixed retarded time.

There are two further requirements that appear to be a necessary for our arguments higher dimensions $d>4$, 
and which seem to go beyond what is usually needed/imposed in $d=4$. The first one, (iv), concerns the
nature of Riemannian metric $s = s_{AB} \, \d x^A \d x^B$ induced on $\Sigma$ by $\tilde g$
(thus $s_{AB}$ is equal to $\gamma_{AB}$ at $r=0=u$), and the value of $\alpha$ on $\I^+$.  If $\M$ is
a spin manifold, as we assume, then $\Sigma$ inherits a spin structure. We require that $(\Sigma,s)$ is a compact and admits a Killing-spinor $\epsilon$ with real
Killing constant. In other words, letting $\D$ be the spin connection on $(\Sigma,s)$, we require that for some $\lambda \in \mr$
\ben\label{killingsp}
\D_X \epsilon = \half i \lambda \, X \cdot \epsilon \ , \qquad \text{for all $X \in T\Sigma$}.
\een
(In this paper, we use the mathematicians' convention denoting Clifford multiplication by $X \cdot$;
in physicists' notation, this would be $X^A \Gamma_A$ in terms of ``gamma-matrices''.)
We then also demand that
\ben\label{alphabndy}
\alpha = \half \lambda^2 \ , \qquad \text{on $\I^+$.}
\een
The value of $\lambda$ can be rescaled by a
corresponding rescaling of $r,u$.
Condition~\eqref{killingsp} implies that $\Sigma$ is an {\em Einstein space} with Ricci tensor
${\rm Ric}_s = \lambda^2 (d-3) \, s$. It turns out (lemma~\ref{lemma1}) that requirement~\eqref{killingsp}
is independent of the chosen cross section $\Sigma$, i.e. if it is satisfied by
one cross section, then it is satisfied for any other.

The second
requirement, (v), is of a global nature. We require that the spacetime $\M$ admits a well defined ``spatial infinity'' and that the
metric functions $\alpha, \beta_A, \gamma_{AB}$ decay like $O(r^{d-3})$
in an open neighborhood of spatial infinity, i.e. for sufficiently
negative $u$ in CGNC's. Such a behavior is
characteristic for stationary solutions such as the Myers-Perry
solution or for the generalized Schwarzschild solution~\eqref{schw}.
Thus, in essence, we ask that, on an asymptotically flat initial data slice, the metric is {\em exactly} equal to a Myers-Perry- or generalized
Schwarzschild metric.
 Both conditions (iv) and (v) are satisfied for
such black holes, with $\Sigma = S^{d-2}$ a standard sphere. 

For future reference, let us summarize our technical assumptions:
\begin{enumerate}
\item[(i)]
$(\M,g)$ is a smooth Lorentzian spin manifold of even dimension $d$.  
There exists a conformal embedding $\M \to \tilde \M$ into a manifold
$\tilde \M$ with boundary $\I = \I^+ \cup \I^-$, and $\I^\pm \cong \RR \times \Sigma$ for a compact, $(d-2)$-dimensional manifold $\Sigma$.

\item[(ii)]
Near $\I^\pm$ the spacetime metric takes the form~\eqref{gform}, for smooth functions $r,u$ and smooth 
tensor fields $\alpha, \beta_A \d x^A, \gamma_{AB} \d x^A \d x^B$ on $\tilde \M$, with $r=0$ being the 
location of $\I^\pm$. 

\item[(iii)]
$(\M,g)$ is a solution to the vacuum Einstein equations ${\rm Ric}_g = 0$.

\item[(iv)] The Riemannian metric $s_{AB} \d x^A \d x^B$ and spin structure induced on $\Sigma$ admits a Killing spinor~\eqref{killingsp}
with real Killing constant $\lambda$, which is related to $\alpha$ by eq.~\eqref{alphabndy}. 

\item[(v)] 
$\M$ admits a well defined ``spatial infinity'' and the
metric functions in~\eqref{gform} decay like $O(r^{d-3})$
in an open neighborhood of spatial infinity. 

\end{enumerate}
A large class of spacetimes satisfying properties (i) to (iii) was shown to exist in~\cite{Anderson:2004pz} (with respect to either $\I^+$ or $\I^-$).
Since property (iv) holds in Minkowski spacetime, the Schwarzschild, or the Myers-Perry spacetimes, and since by results of~\cite{hi}, 
linearized perturbations on such backgrounds affect only higher asymptotic orders of the metric expansion irrelevant for (iv), 
we also expect that it should be satisfied generically for a wide class of solutions. Condition (v) can be viewed as a strengthened version of the notion of an isolated system. 
It is consistent with the well known gluing theorems of~\cite{Corvino:2003sp} (see also~\cite{Chrusciel:2003sr}), showing that arbitrary initial data sets can be modified near spatial infinity so as to coincide e.g. with a Myers-Perry black hole, which in turn have the decay in spatial directions postulated in (v). Thus, in summary we believe that our 
assumptions are reasonable and not overly restrictive.

We next give the definition of Bondi mass, following~\cite{hi}. Let $K_{\tilde g}$ be
the Schouten tensor of $\tilde g$, that is
\ben\label{schouten}
K_{\tilde g} = \tfrac{2}{d-2} \ {\rm Ric}_{\tilde g} - \tfrac{1}{(d-1)(d-2)} \ \tilde g \cdot {\rm Scal}_{\tilde g}  \ ,
\een
and let $C_{\tilde g}$ be the Weyl-tensor of $\tilde g$. For $r>0$ and sufficiently small,
define the ``Bondi mass density'' as
\ben\label{mudef}
\boxed{
\\
\ \  \ \ \mu_{\tilde g} := \tfrac{1}{d-3} \, {r^{-d+4}} \left[ \tfrac{1}{2} \ \Big\langle K_{\tilde g}
- \lambda^2 \tilde g, \   {\rm Hess}_{\tilde g} u \Big\rangle_{\tilde g} - r^{-1} \,
C_{\tilde g}\left( \tfrac{\partial}{\partial r}, {\rm grad}_{\tilde g} r, \tfrac{\partial}{\partial r}, {\rm grad}_{\tilde g} r \right)  \right] \ ,
 \ \ \ \
\\
}
\een
where the gradient vector ${\rm grad}_{\tilde g} r \in T\tilde \M$ is calculated with respect to the unphysical metric $\tilde g$, and similarly the Hessian.
For the motivation/derivation of this expression using Hamiltonian methods see~\cite{hi}, and
for simplicity, we drop the convential factor of $1/8\pi G$.
Let $\Sigma(u,r)$ be the $(d-2)$-dimensional surface of constant $r$ and $u$ defined
near $\I^+$ by our conformal Gaussian null coordinates (CGNC), with
induced integration element $\d S_{\tilde g}$. Our first main
result is

\begin{theorem}\label{thm1}
Under assumptions (i)-(v), the limit
\ben\label{mdef}
m_\Sigma = \lim_{r \to 0} \int_{\Sigma(0,r)} \mu_{\tilde g} \, \d S_{\tilde g}
\een
exists, and defines the Bondi mass of the cross section $\Sigma =\Sigma(0,0)$.
The Bondi news tensor, defined as the limit
\ben
{\rm N} := \lim_{r \to 0} \left[ r^{-d/2+2} (K_{\tilde g} - \lambda^2 \tilde g) \right]
\een
exists on $\I^+$, and we have the mass-loss formula
\ben\label{massloss}
\frac{\d}{\d u} \, m_{\Sigma(u,0)} \Bigg|_{u=0} = - \tfrac{1}{4} \int_{\Sigma} \langle {\rm N}, {\rm N} \rangle_{\tilde g} \, \d S_{\tilde g}
\le 0 \ ,
\een
where $\Sigma(u_0,r_0)=\{u=u_0,r=r_0\}$.
\end{theorem}

\begin{remark}
We remark that the Bondi mass {\em density} $\mu_{\tilde g} \d S_{\tilde g}$ does {\em not} have a well defined limit at $\I^+$, only its {\em integral}. The divergent parts
are, however, shown to be exact forms on $\Sigma(0,r)$, which therefore integrate to zero
by Gauss' theorem. In~\cite{hi}, existence of a finite limit $m_\Sigma$ was demonstrated
as well, but under much more stringent fall-off conditions on the metric.
\end{remark}

 The Bondi-news and mass are shown to have a convenient expression in CGNC's. To this end, we expand all tensors appearing in~\eqref{gform} in powers of $r$, i.e. we consider the
asymptotic expansions
\ben\label{asympt}
\alpha \sim \sum_{n \ge 0} r^n \, \alpha^{(n)} \ , \quad
\beta_A \sim \sum_{n \ge 0} r^n \, \beta^{(n)}_A \ , \quad
\gamma_{AB} \sim \sum_{n \ge 0} r^n \, \gamma^{(n)}_{AB} \ .
\een
The coefficient tensors depend only on $u,x^A$, but not $r$, and by construction $\gamma^{(0)}_{AB} = s_{AB}$ for $u=0$.
\begin{theorem}\label{thm2}
Under assumptions (i)--(v), the Bondi mass is given in CGNC's as
\ben\label{bondim}
m_\Sigma =  (d-2) \int_{\Sigma} \left(\tfrac{1}{8(d-3)}\gamma^{AB(d/2-1)}\dot \gamma_{AB}^{(d/2-1)} - \alpha^{(d-3)}\right) \sqrt{s} \, \d^{d-2} x \ ,
\een
where indices are raised and lowered with $s_{AB}$, and where a `dot' $\cdot$ stands for 
$\partial_u$.
The Bondi news is given in CGNC's as
\ben\label{news}
{\rm N}_{AB} = - \dot \gamma^{(d/2-1)}_{AB} \ ,
\een
with all other components $=0$.
\end{theorem}

\begin{remark}
This formula shows explicitly that the Bondi mass depends on $\alpha^{(d-3)}$, which is
buried inside the asymptotic expansion of the  metric for $d>4$, since the leading
non-trivial coefficient generically turns out to be $\alpha^{(d/2-1)}$. The subtleties of Bondi-energy in higher dimensions are largely due to this circumstance. 

If the spacetime is {\em stationary} with asymptotically timelike Killing field $\xi$, then we may choose our CGNC's such that $\xi = \partial_u$ near 
$\I^+$.  It then follows that all expansion tensors in eq.~\eqref{asympt} are independent of the coordinate $u$, and hence from eq.~\eqref{news} that, as expected, ${\rm N}_{AB} = 0$. 
In particular, the Bondi mass $m_\Sigma$~\eqref{bondim} is now given in terms of $\alpha^{(d-3)}$ alone. 
Going through the proof of lemma~\ref{lemma1}, it also follows that $\alpha^{(d-3)}$ is now the {\em leading} non-trivial metric coefficient in the asymptotic expansions~\eqref{asympt}. 
On the other hand, the ADM mass is given in the stationary case 
by the well-known Komar integral over a cross section tending to spatial infinity, 
\ben
m_{\rm ADM} = \tfrac{d-2}{d-3} \int_\infty \star \d \xi \ .
\een
Working out the integral using~\eqref{gform},~\eqref{asympt} and comparing the result with~\eqref{bondim}
manifestly shows that $m_\Sigma=m_{\rm ADM}$ in the stationary case. 
\end{remark}

\begin{remark}
Theorems~\ref{thm1} and~\ref{thm2} still hold
if~(iv) is replaced by the weaker requirement that $\Sigma$ is an Einstein manifold
with Ricci tensor ${\rm Ric}_s = \lambda^2 (d-3) \, s$, for either real or imaginary
$\lambda$, such as e.g. $\Sigma = S^{d-2}/\Gamma$ or $\Sigma = {\rm H}^{d-2}/\Gamma$, or
$\Sigma = S^{d-2-n} \times S^n$ equipped with a Bohmian Einstein-metric $s_{AB}$~\cite{bohm} ($d=8,10, n \neq 1$) .
\end{remark}

The last, and most important, result of this paper which builds upon the previous two theorems is:

\begin{theorem}\label{thm3}
Suppose that $\M$ admits a smooth spacelike slice $\cS$ intersecting $\I^+$ in the cut $\Sigma$, and such that its (inner) boundaries
$\partial \cS = \cup_i \eH_i$ are comprised of future apparent horizons. If (i)--(v) hold, then $m_\Sigma \ge 0$.
\end{theorem}

This theorem generalizes the known positivity proofs in $d=4$ given by~\cite{Schon:1982re},~\cite{Horowitz:1981uw},~\cite{Ludvigsen:1981gf}, see also~\cite{Chrusciel:2003gn}.
Our proof of theorem~\ref{thm3} works by considering a Witten-spinor on the slice $\cS$, which is why we need to assume a spin-structure on $\M$ in (i).
The investigation of the detailed form of the asymptotic expansion of the metric, i.e. the properties of the coefficients~\eqref{asympt}, and the corresponding expansion of the Witten 
spinor, is an essential ingredient in our proof. The requirement~(iv) of a Killing spinor on $\Sigma$ with {\em real} Killing constant $\lambda$ is needed in order that the corresponding asymptotic expansion of the Witten spinor has the desired properties. For example, for an {\em imaginary} Killing constant, the Ricci curvature ${\rm Ric}_s$ of $\Sigma$ is negative definite. There are
counterexamples to theorem~\ref{thm3} in this case, such as the metric~\eqref{schw} for
$\Sigma = {\rm H}^{d-2}/\Gamma$ and $c<0$. The conformal diagram of that spacetime
can be viewed as the ordinary conformal diagram for Schwarzschild turned by 90~degrees\footnote{
This example was pointed out to us by A. Ishibashi.}.
It has a regular $\I^+$ in our sense, but a negative Bondi-mass $m_\Sigma$.

\begin{remark}\label{remark1}
We finally remark that there is the following characterization of compact Riemannian
manifolds $\Sigma$ admitting a Killing spinor with real Killing constant $\lambda$ in even dimensions:
\begin{itemize}
\item If $\lambda \neq 0$, then it was shown in~\cite{bar} that $\Sigma$
must be a standard sphere $S^{d-2}$, unless $d=8$. In that case $\Sigma$
can also be a nearly K\" ahler Einstein manifold, such that the
warped product $\bar \Sigma=\Sigma \times_{\rho^2} \RR^+$ (cone over $\Sigma$)
has holonomy group ${\rm Hol}(\bar \Sigma) = {\rm G}_2$.
Thus, up to the exception of $d=8$ dimensions, only spherical infinities are possible. In $d=8$, alternatives are (examples taken from~\cite{bar}): (i) $\Sigma = {\rm SU}(3)/{\rm T}^2$ with the
metric inherited from the invariant metric on ${\rm SU}(3)$,
and with ${\rm T}^2$ a maximal torus. (ii) $\Sigma =
({\rm SU}(2) \times {\rm SU}(2) \times {\rm SU}(2))/D({\rm SU}(2))
\cong S^3 \times S^3$, where $D$ is the diagonal subgroup, and the metric is that inherited from the invariant product metric on the
product group. Note that the resulting metric on $S^3 \times S^3$ is not the standard product metric.
Examples of corresponding metrics $g$ on $\M$ satisfying our assumptions in any dimension are the (generalized) Schwarzschild metrics
\ben
\label{schw}
g = -(1-cR^{3-d}) \d t^2 + (1-cR^{3-d})^{-1} \d R^2 + R^2 \ s_{AB} \d x^A \d x^B \ ,
\een
where $s$
is one of the metrics on $\Sigma$ just mentioned with normalization $\lambda=1$ and $c>0$ is proportional to the
Bondi mass (equal to ADM-mass in this case).
One also expects that small non-linear perturbations of such metrics will evolve to
spacetimes satisfying our assumptions.

\item If $\lambda = 0$, then $\Sigma$ admits a parallel spinor. As shown by~\cite{hitchin}, such manifolds can be characterized by their holonomy group. The possibilities are
    ${\rm Hol}(\Sigma) = {\rm SU}(\frac{d-2}{2}), {\rm Spin}(7)$, or $={\rm SL}(\frac{d-2}{2})$ when $d=2$ mod $4$, together with certain semi-direct products
    of these groups by discrete groups, see~\cite{sem}. Conversely, any manifold
    with such a holomomy group admits a parallel Killing spinor.
    We are not aware of any example of an asymptotically flat spacetime with this type of infinity $\Sigma$.
\end{itemize}
\end{remark}

\section{Proofs of theorems~\ref{thm1} and~\ref{thm2}}
\label{sec3}

To prove theorems~\ref{thm1} and~\ref{thm2}, we examine the consequences of Einstein's equations, ${\rm Ric}_g=0$ for the tensor coefficients in the asymptotic expansions~\eqref{asympt} of the tensors
$\alpha,\beta_A,\gamma_{AB}$ on $\Sigma$ appearing in the CGNC form of $g$ given by~\eqref{gform}. The detailed form of Einstein's equations in CGNC's is given
in appendix~\ref{appendix:A}.
As we have already mentioned, the zeroth order coefficient of $\gamma_{AB}$
defines a Riemannian metric $s_{AB}$ on $\Sigma$, which will be used in the following to raise
and lower indices. The Levi-Civita connection of $s_{AB}$ will be called $\D_A$.
The following lemma characterizes the detailed asymptotic form of the metric near $\I^+$.

\begin{lemma}\label{lemma1}
Under the assumptions (i)--(v), the asymptotic expansions~\eqref{asympt} of the tensors
$\alpha,\beta_A,\gamma_{AB}$ on $\I^+$ appearing in the CGNC form~\eqref{gform} of $g$ are restricted by the following.
\begin{itemize}
\item We have
\ben\label{1}
\gamma^{(0)}_{AB} = s^{}_{AB}  \ .
\een

\item We have $\beta_A^{(0)}=0$ and, for  $1 \le n \le \half(d-4)$:
\ben\label{2}
0= \alpha^{(n)}  \ , \quad
0= \beta^{(n)}_A \ , \quad
0= \gamma^{(n)}_{AB} \ .
\een

\item We have for $1 \le n \le d-3$ in the first, and for $1 \le n\le d-4$ in the second equation:
\ben\label{4}
\begin{split}
\beta_A^{(n)} &= -\frac{n}{(n+1)(d-2-n)}\D^B\gamma_{AB}^{(n)} \quad ,\\
\alpha^{(n)}  &=\frac{n-1}{2n(d-3-n)}\D^A\beta_A^{(n)} \quad .
\end{split}
\een

\item We have for $1 \le n \le d-3$ and $d>4$
\ben\label{5}
\gamma^{(n)} \equiv s^{AB}_{} \gamma^{(n)}_{AB} = 0 \quad .
\een


\item
We have
\ben\label{6}
\gamma^{(d-2)} = \frac{3d-10}{8(d-3)} \, \gamma^{(d/2-1)AB}\gamma^{(d/2-1)}_{AB} \ .
\een
\end{itemize}
\end{lemma}

{\em Proof:}  The proof of these relations follows by substitution of the expansions~\eqref{asympt} into the Einstein equations in CGNC as given in appendix~\ref{appendix:A}. We first consider the lowest expansion orders. The equations~\eqref{eq:ei-AB} respectively~\eqref{eq:ei-rA} give at order $r^{-1}$ the relations
\ben
\dot \gamma^{(0)}_{AB} = \beta^{(0)}_A = 0 \ ,
\een
where here and in the following a dot $\cdot$ stands for $\partial_u$.
In particular, $\gamma^{(0)}_{AB}$ is equal to $s_{AB}$ for all $u$, not just $u=0$.
This already completes the proof of eq.~\eqref{2} in $d=4$. In some of the following arguments, 
this special case has to be distinguished, so let us assume $d>4$ for now. 

At order $r^0$, equation~\eqref{eq:ei-ru}
gives the relation
\ben
0=-{\mathcal R}^{(0)} + 2(d-2)(d-3) \, \alpha^{(0)} - (d-3) \, \dot \gamma^{(1)} \ .
\een
Here and in the following $\mathcal{R}_{AB}$ is the Ricci tensor of $\gamma_{AB}$, ${\mathcal R} = \gamma^{AB} {\mathcal R}_{AB}$ the Ricci scalar, 
with a superscript always indicating the coefficients in an asymptotic expansion as in~\eqref{asympt}, 
\ben
{\mathcal R}_{AB} \sim \sum_{n \ge 0} r^n \, {\mathcal R}^{(n)}_{AB} \ , \quad {\mathcal R} \sim \sum_{n \ge 0} r^n \, {\mathcal R}^{(n)} \ .
\een
By assumption (iv), it follows that ${\mathcal R}^{(0)} = \lambda^2(d-2)(d-3)$,
and that $\alpha^{(0)} = \half \lambda^2$, which implies that $\dot \gamma^{(1)} = 0$.\footnote{
It is worth mentioning that the relation $\alpha^{(0)} = \half \lambda^2$, assumed in (iv),
apparently cannot be derived. Indeed, not assuming this relation, we get
$
\alpha^{(0)} = \half \lambda^2
 + \frac{1}{2(d-2)} \dot \gamma^{(1)}
$.
Using the equations~\eqref{eq:ei-uA},~\eqref{eq:ei-rA} and~\eqref{eq:ei-uu} also gives
eq.~\eqref{ba} and
\ben
\D_A\alpha^{(0)} = \dot \beta_A^{(1)} \ , \qquad (d-2) \, \dot\gamma_{AB}^{(1)} = s_{AB} \dot \gamma^{(1)}_{}
\ .
\een
However, these equations do not appear to imply the desired relation, nor $\gamma^{(1)}=0$.}
This equation then implies $\dot \gamma^{(1)} = 0$.
Consider now a sufficiently negative $u$ such that the corresponding cross section $\Sigma(u)$ is near spatial infinity. Then, from assumption (v), it follows that $\gamma^{(1)}=0$ there, hence for all $u$.   At order $r^{0}$, the equation~\eqref{eq:ei-AB} next gives
\ben
0 = (d - 1)(d - 4) \dot
\gamma^{(1)}_{AB} + 2(d - 1)\, {\mathcal R}^{(0)}_{AB} -
s_{AB}^{} {\mathcal R}^{(0)} - 2d(d - 3)\, \alpha^{(0)} s_{AB}  \ .
\een
Using what we have found so far and using assumption (iv), we
get that $\dot \gamma^{(1)}_{AB}=0$. From assumption (v),
$\gamma^{(1)}_{AB}=0$ for sufficiently negative $u$, hence everywhere.
It then follows that ${\mathcal R}^{(1)}=0$ and from equation~\eqref{eq:ei-rr} that $\gamma^{(2)}=\gamma^{(3)}=0$.
We next consider, at order
$r^{0}$, equation~\eqref{eq:ei-rA}, giving
\ben\label{ba}
\beta^{(1)}_A = -\frac{1}{2(d - 3)} \, ( \D^B \gamma^{(1)}_{AB}-\D_A^{} \gamma^{(1)}_{}) = 0 \ .
\een
The argument proceeds in a similar pattern by showing
next $\alpha^{(1)}=0$.
The idea is to use systematically increasing orders $r^n$ of the Einstein equations
presented in the appendix~\ref{appendix:A}.
First, for $n < d/2-1$ one assumes inductively
\ben\label{induct}
\begin{split}
0= \alpha^{(k)}_{} = \gamma_{AB}^{(k)} & \quad \text{for $1 \le k < n$}\\
0= \beta^{(k)}_A & \quad \text{for $0\le k < n$}.
\end{split}
\een
It follows that
\ben
0= \gamma^{(k)} \quad \text{for $1\le k <2n$} 
\een
from eq.~\eqref{eq:ei-rr}, and it also follows that
\ben
0= {\mathcal R}^{(k)}_{AB}  \ , \quad 0={\mathcal R}^{(k)}  \quad \text{for $1 \le k < n$} \ ,
\een
and then from~\eqref{eq:ei-AB} that
\ben
0=(d-1)(d-2-2n) \, \dot \gamma^{(n)}_{AB} \ .
\een
This gives $\gamma^{(n)}_{AB}=0$ as long as $n < d/2-1$, by the same argument as above.
The inductive assumption also gives, for $m<2n-1$,
\ben\label{mind}
\begin{split}
0=&-{\mathcal R}^{(m)} + 2(m+1)(d-2)(d-m-3) \, \alpha^{(m)} - (m+1)(d-3) \, \D^A \beta^{(m)}_A \ , \\
0=&d \, {\mathcal R}^{(m)} + 2(d-2)(d+m)(d-m-3) \, \alpha^{(m)} - 2[d(d-3)-m+1] \, \D^A \beta^{(m)}_A \ ,\\
0=&(d-2-m)(m+1)\, \beta^{(m)}_A + m \, \D^B \gamma^{(m)}_{AB} \ ,
\\
0=&2(d-3-m) \, \D_A\alpha^{(m-1)} + (m+1) \, \dot\beta_A^{(m)} +
\D^B_{}\dot\gamma_{AB}^{(m)} - \D^B_{} \D_A \beta_B^{(m)} + \\
& \D^B_{} \D_B \beta_A^{(m)} - 2(m-2)(d-3-m) \, \alpha^{(0)}_{}\beta_A^{(m-1)}
\end{split}
\een
from equations~\eqref{eq:ei-ru}, the trace of~\eqref{eq:ei-AB},~\eqref{eq:ei-rA},
and~\eqref{eq:ei-uA}, respectively.
The induction step for the remaining
quantities in eq.~\eqref{induct} now follows from~\eqref{mind}, demonstrating
eqs.~\eqref{2},~\eqref{5}. The relations~\eqref{4} follow again from~\eqref{mind}.
Equation~\eqref{6} follows from~\eqref{eq:ei-rr}. The case $d=4$ is similar. For further
details, we refer to~\cite{thorne}. \qed

The lemma and also its proof shows how the expansion coefficients are successively determined 
by those at the lowest orders, which in turn are basically fixed by our assumptions. However, this 
process does not continue without limit, and at definite expansion orders, the expansion coefficients remain undetermined. This
happens precisely for $\alpha^{(d-3)}, \beta^{(d-2)}_A$, and for $\gamma^{(d/2-1)}_{AB}$, which will then feed into the higher orders. These
coefficients therefore represent invariants of the solution under consideration. Not surprisingly, the first and last enter the Bondi mass and 
news according to theorem~\ref{thm2}, whereas the second enters
the Bondi angular momentum, which we do not consider in this paper.

To prove theorems~\ref{thm1} and~\ref{thm2}, we work out
the Schouten tensor $K_{\tilde g}$, the Weyl tensor
$C_{\tilde g}$, and the Hessian ${\rm Hess}_{\tilde g} u$
in terms of the unphysical metric $\tilde g$, see ~\eqref{gform}, and use eqs.~\eqref{1},~\eqref{2} and~\eqref{5} from the lemma. We then substitute those results into the expression for the Bondi mass density~\eqref{mudef} and expand the result in
powers of $r$ as:
\ben
\mu \sim r^{-d+3} \sum_{n \ge 0} r^n \, \mu^{(n)} \ .
\een
A lengthy calculation shows $\mu^{(0)}=0$ and
\ben
\mu^{(n)}  =
-\frac{n(n+1)}{d-3} \, \alpha^{(n)} =
-\frac{(n+1)(n-1)}{2(d-3-n)(d-3)} \, \D^A\beta_A^{(n)}
\qquad \text{for $1 \le n < d-3$,}
\een
whereas
\ben
\mu^{(d-3)}=
 -(d-2) \, \alpha^{(d-3)} + \frac{d-2}{8(d-3)} \, \gamma^{(d/2-1)}_{AB} \dot \gamma^{(d/2-1)AB} \qquad .
\een
To get to the second expression in  $\mu^{(n)}$ we used~\eqref{4}, and in the
calculation of $\mu^{(d-3)}$ we have used eq.~\eqref{7}. Using
eq.~\eqref{5}, the integration element on the surfaces $\Sigma(u,r)$ of constant
$r,u$ is seen to behave as $\d S_{\tilde g} = \sqrt{s} \, \d^{d-2} x$ plus
terms of order $r^{d-2}$. Substituting these results into the formula~\eqref{mdef}
for the Bondi-mass $m_\Sigma$, we find
\ben
\begin{split}
m_\Sigma &= \lim_{r \to 0}  \sum_{n=1}^{d-3} r^{n-d+3} \int_\Sigma \mu^{(n)} \sqrt{s} \, \d^{d-2} x\\
&= \int_\Sigma \mu^{(d-3)} \sqrt{s} \, \d^{d-2} x \ ,
\end{split}
\een
because all potentially divergent terms as $r \to 0$, namely those with $n<d-3$, are in fact integrals of total divergences, and hence fortunately vanish. Hence, the limit~\eqref{mdef} defining $m_\Sigma$ indeed exists. This proves the first statement in theorem~\ref{thm1}. The finite piece coming from $\mu^{(d-3)}$
gives exactly the formula~\eqref{bondim} for $m_\Sigma$ stated in theorem~\ref{thm2}. Equation~\eqref{news} for the Bondi-news tensor follows from a similar calculation, whereas to demonstrate the mass-loss formula~\eqref{massloss}, we must also use  $u$-derivatives of~\eqref{6}. For details, we refer to~\cite{thorne}.

\section{Proof of theorem~\ref{thm3}}

Our proof of that theorem is based on spinor methods, so it is essential to assume
that $\M$ carries a fixed spin-structure. Let ${\rm Cliff}(T\M)$ be the corresponding Clifford bundle associated with the Clifford algebra ${\rm Cliff}_{d-1,1}$ of the quadratic form $-x_0^2 + x_1^2 + \dots + x_{d-1}^2$ on $\RR^d$.
Relative to a local basis $e_\mu, \mu=0, \dots, d-1$ of $T_x\M$,
${\rm Cliff}(T_x \M)$ consists of the identity $I$ and the expressions $e_{\mu_1} \cdot \dots \cdot e_{\mu_p}$ subject to the relations
\ben
e_\mu \cdot e_\nu + e_\nu \cdot e_\mu = 2 g(e_\mu,e_\nu) \, I \ .
\een
Spinors $\psi$ are smooth sections in the complex vector bundle $\$ $ associated with
the complex $2^{d/2}$-dimensional fundamental representation of the complexified
Clifford algebra $\CC{\rm liff}_{d-1,1}= \CC \otimes {\rm Cliff}_{d-1,1}$. The standard
anti-linear automorphism of the Clifford algebra $\CC{\rm liff}_{d-1,1}$ gives rise to a sesquilinear
inner product between spinors which can be lifted to $\$ $ and used to identify spinors $\psi \in \$ $
with complex anti-linear maps $\overline \psi: \$ \to \CC$. The spin connection
is denoted $\nabla$. Using this, we define a 2-form $Q$ on $\M$ by:
\ben\label{Qdef}
Q(X,Y) = {\rm Re} \left( \overline \psi \ Y \cdot \nabla_X \psi - \overline \psi \ X \cdot \nabla_Y \psi \right) \ ,
\een
for all $X,Y \in T\M$. We now assume that there exists a spacelike $(d-1)$-dimensional smooth submanifold $\cS$ of
$\M$ such that its closure  $\tilde \cS$  in the unphysical spacetime $\tilde \M$ meets $\I^+$ transversally in the 
cut $\Sigma$, which is in this sense an ``outer boundary'' of $\cS$. With the application to black hole spacetimes in mind, 
 we also allow $\cS$ to have one or more ``inner boundaries'', $\eH_i$, each of which is assumed to be a (future) apparent horizon.  To simplify our
subsequent calculations we define the surface $\cS$ as
\ben
\cS = \{ u-\half r = 0\} \ ,
\een
near $\I^+$, so that the cut $\Sigma$ of $\I^+$ meeting $\cS$ is at the value $u=0=r$ in CGNC's.
In Minkowski space, our choice of $\cS$  corresponds, asymptotically, to a hyperboloidal surface
$t=(1+\sum x_i^2)^{1/2}$, as can be seen explicitly from the relationship $r=(\sum x_i^2)^{-1/2},\ t=u+r^{-1}$ between CGNC and Cartesian coordinates in Minkowski space. Except possibly for certain special cases, this is true more generally under our assumptions (i)--(v)
and the vacuum Einstein equations. Indeed, we have already noted in remark~\ref{remark1} that $\Sigma = S^{d-2}$
is necessarily a round sphere, unless $\lambda=0$ or $d = 8$. Except in that case, it follows from lemma~\ref{lemma1} that the induced metric on $\cS$  approaches that
of hyperbolic space ${\rm H}^{d-1}$. This can be seen explicitly simply by eliminating $\d u$ in eq.~\eqref{gform}
using $\d u = \half \d r$ on $\cS$.

Now let $e_0, e_1, \dots, e_{d-1}$ be a positively oriented orthonormal frame adapted to $\cS$, in such
a way that $e_0$ is the unit future timelike normal, and $e_1,\dots,e_{d-1}$ is a positively
oriented orthonormal basis of $T\cS$ at each point. Furthermore,
suppose that $\psi$ is a solution to the pair of equations,
\ben\label{witten}
0= \sum_{i=1}^{d-1} e_i \cdot \nabla_{e_i} \psi  \ , \qquad
0= \nabla_{e_0} \psi \ , \qquad
\text{on $\cS$.}
\een
The first equation is in this context called the Witten equation. The second equation
can be viewed as prescribing how $\psi$ is extended, to first order, off $\cS$. In fact,
we will only need both equations on $\cS$. The Bochner-type argument due
to~\cite{witten} shows that $\d \star Q$ is, on $\cS$, in the same orientation
class as $e_1 \wedge \dots \wedge e_{d-1}$. Therefore, for any {\em compact} subset $\cK \subset \cS$,
we get from Stokes theorem the inequality
\ben
0 \le \int_{\cK} \d \star \! Q = \int_{\partial \cK} \star Q \ .
\een
The idea is, as usual, to use this identity for the surface $\cK = \cS$, but this is not directly possible, as
$\cS$ is non-compact in $\M$, and the integral looks divergent at first sight. So let us consider instead the part of $\cS(r_0) \subset \cS$ having
$r\ge r_0>0$, i.e. which is bounded away from $\I^+$. The boundary then consists of
$\partial \cS(r_0) = \Sigma(r_0) \cup (\cup_i -\eH_i)$, where $\Sigma(r_0) =\{ u=\half r_0 , r=r_0 \}
\equiv \Sigma(\half r_0,r_0)$ in our earlier notation. Additionally, let us impose on $\psi$ the elliptic boundary conditions 
at the inner boundaries $\eH_i$ pointed out in~\cite{horowitz}, i.e. 
if $e_1$ denotes the outward pointing normal of $\eH_i$ within $\cS$,  we impose
\ben\label{innerbc}
\half (e_1 \wedge e_0) \cdot \psi = \psi \quad \text{on each $\eH_i$.}
\een
Then, as shown in~\cite{horowitz}, the corresponding contribution to the boundary integral vanishes, and we have for $r>0$
(and sufficientlly small)
\ben\label{pos}
0 \le \int_{\Sigma(r)} \star Q = r^{-d+2} \int_{\Sigma(r)} Q(\tilde e_+, \tilde e_-) \, \d S_{\tilde g} \ .
\een
In the last expression, we have used the bi-normal
$\tilde e_+ \wedge \tilde e_-$
(relative to $\tilde g$) to $\Sigma(r)$
given by
\ben\label{epm}
\tilde e_+ = \frac{\partial}{\partial r} \ , \qquad
\tilde e_- = r^2\alpha \frac{\partial}{\partial r} + \frac{\partial}{\partial u} \ ,
\een
verifying $\tilde g(\tilde e_+, \tilde e_-)=1, \tilde g(\tilde e_\pm, \tilde e_\pm) = 0$.
The idea is now to try to show that the desired solution to the Witten equation indeed exists on $\cS$, 
that the expression on the right side of \eqref{pos} remains finite as $r \to 0$, and that it tends, in fact, 
to the Bondi mass $m_\Sigma$ as $r \to 0$. This would then evidently show that $m_\Sigma \ge 0$, 
and thereby complete the proof of theorem~\ref{thm3}. As we briefly will argue in the end, 
the existence proof of the Witten spinor $\psi$ is not substantially different from 
$d=4$ dimensions, where the corresponding statement was shown e.g. in~\cite{Chrusciel:2003qr}.   
However, the proof that \eqref{pos} remains finite, and that it tends to the expression for the 
Bondi mass $m_\Sigma$ as $r \to 0$ is rather more complicated in $d>4$ dimensions, and will therefore occupy 
most of the remainder of this section.

To be able to make progress, we need to know in precise detail the asymptotic expansion
of $\psi$ on $\cS$ near $r=0$, i.e. near $\Sigma$. These calculations are best performed
in terms of the unphysical metric $\tilde g$, see \eqref{gform}, with associated spin structure ${\rm Cliff}(T \tilde \M)$
and spin-connection $\tilde \nabla$. The natural isomorphism ${\rm Cliff}(T \tilde \M) \to {\rm Cliff}(T \M)$ defines a corresponding map $\tilde \$ \to \$ $. A spinor field is said to
be smooth at $\I^+$ if the spinor field in the unphysical bundle $\tilde \$ \to \tilde \M$
obtained by this map can be smoothly extended across $\I^+$. We now suppose that a solution
to~\eqref{witten} exists such that $r^{1/2} \psi = \tilde \psi$ is smooth at $\I^+$, which will be justified below. Then we
may write, on $\cS$ near $\I^+$:
\ben\label{asympt1}
\psi \sim r^{-\half} \sum_{n \ge 0} r^n \ \psi^{(n)} \ ,
\een
where each $\psi^{(n)}$ is smooth and near $\I^+$ and satisfies $\tilde \nabla_{\partial/\partial r} \psi^{(n)} = 0$. Since the $\psi^{(n)}$ are parallel transported in the $r$-direction near $\I^+$, we can identify their restriction to $\Sigma(r)$ with a spinor on $\Sigma=\Sigma(0)$ via
parallel transport. The Clifford elements ${\sf P}_\mp = \half \tilde e_{\mp} \cdot \tilde e_{\pm} \in {\rm Cliff}(T\tilde \M)$ are projections satisfying ${\sf P}_+ {\sf P}_- =
{\sf P}_- {\sf P}_+ = 0$. The split $\$ = \$_+ \oplus \$_-$ into invariant
subspaces $\$_\pm = {\sf P}_\pm \$ $ of complex dimension $2^{d/2-1}$ corresponds to the decomposition of the restriction of ${\rm Cliff}(T \tilde \M)$ to $T\Sigma$ as ${\rm Cliff}(T\Sigma) \oplus {\rm Cliff}(T\Sigma)$, so that $\$_{\pm \ x}$ are ismorphic modules of $\CC{\rm liff}(T_x \Sigma)$ for each $x \in \Sigma$. In terms of this split, we may represent the generators of
${\rm Cliff}(T\tilde \M)$ as
\ben
\tilde e_+ =
\sqrt{2} \left(
\begin{matrix}
0 & I \\
0 & 0
\end{matrix}
\right) \ , \quad
\tilde e_- =
\sqrt{2} \left(
\begin{matrix}
0 & 0\\
I & 0
\end{matrix}
\right) \ , \quad
\tilde e_{A} = \left(
\begin{matrix}
\Gamma_A & 0 \\
0 & -\Gamma_A
\end{matrix}
\right) \ ,
\een
where $\tilde e_A \in T\Sigma$, and where $\Gamma_A$ are the corresponding
generators of ${\rm Cliff}(T\Sigma)$:
\ben
\Gamma_A \cdot \Gamma_B + \Gamma_B \cdot \Gamma_A = 2 \, s_{AB} \, I_{\$_\pm} \ .
\een
The intrinsic spin-connection associated with ${\rm Cliff}(T\Sigma)$ is
called $\D$. After these preliminaries, we can state the following lemma:

\begin{lemma}\label{lemma2}
Assume that $\psi$ is smooth and satisfies~\eqref{witten} on $\cS$ near $\Sigma$, with
asymptotic expansion~\eqref{asympt1}. Write $\psi_\pm^{(n)} = {\sf P}_\pm \psi^{(n)}$,
and assume $\psi_+^{(0)} = \epsilon$, where $\epsilon$ is a Killing spinor on $(\Sigma,s)$, and $\psi^{(0)}_- = 0$. We have, for\footnote{If the 
order of a coefficient is negative (such as $\alpha^{(n-2)}$ for $n=1$), then that coefficient is by convention 
set to 0.} $1 \le n<d-1$:
\ben\label{spinrec}
\begin{split}
\psi_-^{(n)} = & \frac{1}{d/2-n}\Big[
\frac{1}{\sqrt{2}} \, \Gamma^A \D_A \psi_+^{(n-1)} +
\frac{2n-d}{8\sqrt{2}} \Gamma^A \beta^{(n-1)}_A \psi_+^{(0)} -
\frac{1}{4\sqrt{2}} (\D^B \gamma^{(n-1)}_{AB} - \D^{}_A \gamma^{(n-1)} ) \psi_+^{(0)} \Big] \\
\psi_+^{(n)} = & \frac{-1}{2n}\Big[
-\frac{1}{\sqrt{2}} \, \Gamma^A_{} \D_A \psi^{(n-1)}_- -
(\half d + n -3) \, \alpha^{(0)}_{} \psi^{(n-2)}_+ +
\frac{d-9}{\sqrt{2}} \, \Gamma^A_{} \beta_A^{(n-2)} \psi^{(1)}_- + \\
& \hspace{1cm}
\frac{n-4}{2} \, \alpha^{(0)}_{} \gamma^{(n-2)}_{} \psi^{(0)}_+ +
\frac{1}{4} \, \D^{}_B \beta^{(n-2)}_A \Gamma^A \Gamma^B \psi^{(0)}_+ +
\frac{1}{4} \, \dot \gamma^{(n-1)}_{} \psi^{(0)}_+ - \\
& \hspace{1cm}
\frac{d-2}{2} \, \alpha^{(n-2)}_{} \psi^{(0)}_+ -
\frac{1}{4\sqrt{2}} (\D^B \gamma^{(n-2)}_{AB} - \D^{}_A \gamma^{(n-2)} ) \Gamma^A \psi^{(0)}_+
\Big] \ . 
\end{split}
\een
The first relation does not hold for $n=d/2$, where the expression in brackets necessarily has to vanish.
For $n=d-1$, the second relation is instead:
\ben
\begin{split}
\psi_+^{(d-1)} = & \frac{-1}{2(d-1)}\Big[
-\frac{1}{\sqrt{2}} \, \Gamma^A_{} \D_A \psi^{(d-2)}_- -
(\frac{3}{2} d -4) \, \alpha^{(0)}_{} \psi^{(d-3)}_+ +
\frac{d-9}{\sqrt{2}} \, \Gamma^A_{} \beta_A^{(d-3)} \psi^{(1)}_- + \\
& \hspace{1cm}
\frac{d-5}{2} \, \alpha^{(0)}_{} \gamma^{(d-3)}_{} \psi^{(0)}_+ +
\frac{1}{4} \, \D^{}_B \beta^{(d-3)}_A \Gamma^A \Gamma^B \psi^{(0)}_+ +
\frac{1}{4} \, \dot \gamma^{(d-2)}_{} \psi^{(0)}_+ - \\
& \hspace{1cm}
\frac{d-2}{2} \, \alpha^{(d-3)}_{} \psi^{(0)}_+ -
\frac{1}{4\sqrt{2}} (\D^B \gamma^{(d-3)}_{AB} - \D^{}_A \gamma^{(d-3)} ) \Gamma^A \psi^{(0)}_+ -\\
&\hspace{1cm}
\frac{1}{4} \, \gamma^{(d/2-1)AB}_{} \dot \gamma_{AB}^{(d/2-1)} \psi^{(0)}_+
\Big] \ .
\end{split}
\een
All indices are raised and lowered with $s_{AB}$ and a `dot' $\cdot$ stands for $\partial_u$. 
\end{lemma}

{\em Proof:} The relations basically follow from the Witten equation~\eqref{witten}. They are derived most easily working in
CGNC's~\eqref{gform}, and working in a local frame defined by the null tetrad for $\tilde g$ consisting of $\tilde e_\pm$ (see eq.~\eqref{epm})
complemented with an orthonormal frame $\tilde e_1, \dots, \tilde e_{d-2}$ perpendicular to the span of $\tilde e_\pm$. We may choose
$r,x^A$ as coordinates on $\cS$ near $\I^+$, and express $u=\half r$. Expressing the normal to $\cS$ in terms of CGNC's, the
second equation in~\eqref{witten} then gives\footnote{In this, and the following equation, we deviate from the usual
convention in place in the body of the paper in that $\gamma^{AB}$ is the inverse of $\gamma_{AB}$, rather
than indices raised by $s^{AB}$, as usual.} on $\cS$:
\ben
0=\Big[ (-1+r^2\alpha + \half r^2 \gamma^{AB}\beta_A \beta_B) \nabla_r + \half \nabla_u + \beta^A \nabla_A \Big] \psi \ .
\een
This is used to express any $u$-derivative by $r$- and $x^A$-derivatives in the first equation in~\eqref{witten}.
One thereby finds the equation
\ben
\begin{split}
0 =& \tilde e_+ \cdot \tilde \nabla_r \tilde \psi
+ (2-r^2\alpha-r^2\gamma_{AB} \beta^A\beta^B) \, \tilde e_-  \cdot \tilde \nabla_r \tilde \psi
+ r\beta^A\tilde e_A \cdot \tilde \nabla_r\tilde\psi -
r\beta^A\, \tilde e_- \cdot \tilde \nabla_A \tilde \psi + \gamma^{AB} \, \tilde e_A \cdot \tilde\nabla_B \tilde\psi \\
& - \frac{d}{2}r^{-1}\, \tilde e_+\cdot \tilde\psi - \frac{d-2}{2}r\alpha\, \tilde e_- \cdot \tilde\psi - \frac{d}{2}\beta^A\tilde e_A \cdot \tilde\psi \
\end{split}
\een
for the spinor $\tilde \psi = r^{1/2} \psi$.
As usual, a dot like in $\tilde e_\pm \cdot \tilde \psi$ denotes Clifford multiplication in ${\rm Cliff}(T\tilde\M)$, and $\tilde \nabla$ is
the associated spin-connection of the unphysical metric $\tilde g$. Now we substitute
the expansion~\eqref{asympt1}. Then we apply ${\sf P}_\pm \tilde \nabla^n_r$, take $r=0$, use $\tilde \nabla_r\psi^{(n)} = 0$, and use the
Killing spinor equation~\eqref{killingsp} as well as identities in lemma~\ref{lemma1}. This then leads to the recursive
formulas after a rather lengthy calculation, which we omit.
\qed

The spinorial `coefficients' $\psi^{(n)}{}_\pm$ for all $n$ are uniquely determined by the pair of spinors $\psi^{(0)}{}_+, \psi^{(d/2)}{}_-$ intrinisic to $\Sigma$, and
the local geometry near $\I^+$ encoded in the expansion coefficients~\eqref{asympt} of the metric in CGNC's.
By assumption, the first spinor is equal to $\psi^{(0)}{}_+=\epsilon$. By contrast, the second spinor $\psi^{(d/2)}{}_-$ cannot be expressed in terms of $\epsilon$ and/or the local
geometry near $\I^+$, but is instead determined by globally solving the Witten equation.

It follows from the recursive formulas of lemma~\ref{lemma2} that up to order $d/2$, most expansion coefficients
are zero:
\ben
\label{firstterms}
\begin{split}
&0=\psi_+^{(1)} = \psi_+^{(d/2)} \ , \qquad
\psi_-^{(1)} = \tfrac{1}{\sqrt{2}} i\lambda \epsilon \ , \\
&0 = \psi^{(2)}_\pm = \psi^{(3)}_\pm=\cdots = \psi^{(d/2-1)}_\pm \  .
\end{split}
\een
These results, and the recursive equations of lemma~\ref{lemma2} are now used to determine the coefficient functions on $\Sigma$ in the expansion of
\ben\label{Qexp}
Q(\tilde e_+, \tilde e_-) \sim \sum_{n \ge 0} r^n \ Q^{(n)} \ ,
\een
on $\cS$ near $r=0$.
Inserting the exansion~\eqref{asympt1} into the definition of $Q$, eq.~\eqref{Qdef}, and using also~\eqref{epm}, and~\eqref{witten}, one first shows that
$Q^{(0)}=0$ and, for $1 \le n \le d-2$,
\ben\label{Qndef}
Q^{(n)} = 2 \ {\rm Re} \Big[
\tfrac{1}{\sqrt{2}} (n-1) \ \langle \psi^{(1)}_-,  \psi^{(n)}_- \rangle + \sqrt{2} (n+1) \ \langle \psi^{(0)}_+ ,\psi^{(n+1)}_+ \rangle
- \tfrac{1}{\sqrt{2}} (n-1) \alpha^{(0)}_{} \langle \psi^{(0)}_+, \psi^{(n-1)}_+ \rangle
\Big] \ .
\een
Here, $\langle \ \ , \ \ \rangle$ is the positive definite hermitian inner product on $\$_\pm$, the spinor bundle over $\Sigma$, that is
induced by the usual anti-linear automorphism of $\CC{\rm liff}(T\Sigma)$. To show the limit of the expression~\eqref{pos}
as $r \to 0$ exists, and to understand its nature, we must consider potentially divergent and finite terms. In view
of eq.~\eqref{Qexp}, these can come from any $Q^{(n)}$ having $n < d-2$ and $n=d-2$, respectively. However, the spinors $\psi^{(n)}{}_\pm$ are known to vanish only up to $n < d/2$, but are generally
non-zero for $n \ge d/2$. Worse still, they are not determined by the local geometry near $\I^+$, and hence essentially unknown. Equation~\eqref{Qndef} therefore appears to offer little hope of
progress at first sight. Fortunately, it helps at this stage to
use the recursion relations provided by lemma~\ref{lemma2} for $n \ge
d/2$. Indeed, let us substitute the second recursion relation eq.~\eqref{spinrec} for in the second term in~\eqref{Qndef}. In the last
term, we use the Killing-spinor equation~\eqref{killingsp} in the form
\ben
\Gamma^A \Gamma^B \D_A \D_B \, \psi_+^{(0)} = -\half (d-2)^2 \, \alpha^{(0)}_{} \psi^{(0)}_+ \ ,
\een
move one derivative on the second factor at the expense of
a total divergence, and apply the first recursion relation in eq.~\eqref{spinrec} from lemma~\ref{lemma2}. It turns out that this leads to rather non-trivial cancelations between various terms. A lengthy calculation shows that the final result can be written as, for $1\le n<d-2$
\ben
Q^{(n)} = \sqrt{2} \Big[
\tfrac{1}{2} \, \alpha^{(0)}_{} \gamma^{(n-1)} + \tfrac{1}{4} \D^A_{} \beta^{(n-1)}_A - (\tfrac{1}{4} \dot \gamma^{(n)}_{} - \half(d-2) \alpha^{(n-1)}_{})
\Big] |\epsilon|^2 + \D^A_{} w_A^{(n)}
\een
 where the precise form of the last divergence term is unimportant for us. Furthermore, when $n=d-2$, we find instead:
\ben
\begin{split}
Q^{(d-2)} =&  \sqrt{2} \Big[
\tfrac{1}{2} \, \alpha^{(0)}_{} \gamma^{(d-3)}_{} + \tfrac{1}{4} \D^A_{} \beta^{(d-3)}_A - \\
& \hspace{1cm} (\tfrac{1}{4} \dot \gamma^{(d-2)}_{} - \half(d-2) \alpha^{(d-3)}_{}
- \tfrac{1}{4} \gamma^{(d/2-1)}_{AB} \dot \gamma^{(d/2-1)AB}_{}
)
\Big] |\epsilon|^2
+ \D^A_{} w_A^{(d-2)} \ .
\end{split}
\een
We note that, fortunately and rather non-trivially, the unknown spinors $\psi^{(n)}{}_\pm, n \ge d/2$ have now
completely dropped out.
We next use that, since $\epsilon$ is a Killing spinor on $\Sigma$ with real Killing constant $\lambda$, see~\eqref{killingsp}, the quantity $\langle \epsilon, \epsilon \rangle =|\epsilon|^2$ is
constant\footnote{This would be false for imaginary Killing
constant $\lambda$.} and positive on $\Sigma$. Then, we use the results of lemma~\ref{lemma1} that $\alpha^{(n-1)}=\gamma^{(n)}=\gamma^{(n-1)}=0$ up to a total divergence in the
range $1 \le n < d-2$. It follows immediately that $Q^{(n)}$ is also a total divergence on
$\Sigma$, i.e.
\ben
Q^{(n)} =  \D^A_{} w_A^{(n)} \qquad \text{for $1\le n < d-2$,}
\een
for a new $w_A^{(n)}$.  Furthermore, using eq.~\eqref{6}, it is found for $n=d-2$ that
\ben\label{qdm2}
Q^{(d-2)} = \tfrac{1}{\sqrt{2}} (d-2) \Big[ -\alpha^{(d-3)}_{}
+ \tfrac{1}{8(d-3)} \gamma^{(d/2-1)}_{AB} \dot \gamma^{(d/2-1)AB}_{}
\Big] |\epsilon|^2 + \D^A_{} w_A^{(d-2)} \ ,
\een
for a new $w_A^{(d-2)}$. Thus, because total divergences integrate to zero over the closed compact surface $\Sigma$, it follows that
\ben
\begin{split}
\lim_{r \to 0}  \left[ r^{-d+2} \int_{\Sigma(r)} Q(\tilde e_+, \tilde e_-) \, \d S_{\tilde g} \right] &= \lim_{r \to 0}
\sum_{n = 1}^{d-2} r^{-d+2+n} \int_{\Sigma}  Q^{(n)} \, \sqrt{s} \, \d^{d-2} x \\
&=
\int_\Sigma Q^{(d-2)} \, \sqrt{s} \, \d^{d-2} x \ .
\end{split}
\een
We also used that $\d S_{\tilde g} = \sqrt{s} \, \d^{d-2} x + O(r^{d-2})$, which follows from lemma~\ref{lemma1}.
Therefore, in view of~\eqref{qdm2}, it follows that, choosing the normalization $|\epsilon|^2=\sqrt{2}$:
\ben
\lim_{r \to 0}  \left[ r^{-d+2} \int_{\Sigma(r)} Q(\tilde e_+, \tilde e_-) \, \d S_{\tilde g} \right]
= (d-2) \int_\Sigma
\left(\tfrac{1}{8(d-3)}\gamma^{AB(d/2-1)}\dot\gamma_{AB}^{(d/2-1)} - \alpha^{(d-3)}\right) \sqrt{s} \, \d^{d-2} x \ . \non
\een
In particular, we have demonstrated the rather non-obvious fact that the limit actually exists. (Note that the limit of the {\em integrand}
on the left side does {\em not} exist.)
By theorem~\ref{thm2}, the right side is equal to the Bondi-mass $m_\Sigma$. Hence we have shown:

\begin{lemma}\label{lemma3}
Under the same hypothesis as in lemma~\ref{lemma2}, it follows that
\ben
m_\Sigma = \lim_{r \to 0}  \left[ r^{-d+2} \int_{\Sigma(r)} Q(\tilde e_+, \tilde e_-) \, \d S_{\tilde g} \right] \ ,
\een
where $m_\Sigma$ is the Bondi mass. 
\end{lemma}

In particular, if there exists smooth slice $\cS$ stretching between apparent horizons $\eH_i$ and the cut $\Sigma$ of $\I^+$ as we are assuming, and if there exists a smooth 
spinor field satisfying the Witten equation~\eqref{witten}, 
with asymptotic expansion~\eqref{asympt1}, with $\psi^{(0)}{}_+ = \epsilon, \psi^{(0)}{}_-=0$, and with the boundary conditions at $\eH_i$ as in~\cite{horowitz}, 
then~\eqref{pos} holds true and lemma~\ref{lemma3} immediately tells us that $m_\Sigma \ge 0$, thereby completing the proof of theorem~\ref{thm3}. 

We will not try to investigate conditions under which a slice $\cS$ of the desired type exists, and therefore leave it as an assumption as in the statement of theorem~\ref{thm3}. 
But we need to establish the existence of a spinor field with the desired properties. 
The proof of existence is not substantially altered in higher dimensions compared to the case $d=4$
treated in~\cite{Chrusciel:2003gn,Chrusciel:2003qr}, so we are brief. One can argue as follows. First, define, using the recursive
formulas of lemma~\ref{lemma2}, the spinors $\chi^{(n)}{}_\pm$ such that $\chi^{(0)}{}_-=0$, and
\ben
\chi^{(0)}_+ = \epsilon \ , \qquad \chi^{(d/2)}_-=0 \ .
\een
As we have mentioned, these conditions are sufficient to generate the complete asymptotic expansion $\chi^{(n)}{}_\pm$ for all $n$, satisfying the analog of~\eqref{firstterms}. Let $\chi$ be a spinor with smooth
extension $\tilde \chi = r^{1/2} \chi$ to $\tilde \cS$, having
these prescribed expansion coefficients as in eq.~\eqref{asympt1}, and vanishing on the ``inner boundaries''. This spinor
field will not, in general, satisfy the Witten equation~\eqref{witten}. But because its asymtptotic expansion coefficients 
have been constructed obeying the same recursion relations as a spinor which does (this is the content of lemma~\ref{lemma2}), it is shown to satisfy the Witten equation with a source whose asymptotic expansion coefficients vanish up to some order depending on $d$. In fact, 
using also lemma~\ref{lemma1}, it is seen that the source is square integrable on
the asymptotically hyperbolic manifold $\cS$. By standard
arguments, it is then possible to define the desired spinor as $\psi = \chi + \delta \psi$, where $-\delta \psi$ is the unique $L^2$-solution to the equation with source
and inner boundary conditions as in eq.~\eqref{innerbc}, see e.g.~\cite{Chrusciel:2003qr}. Furthermore,
using e.g. the results and methods of~\cite{melrose}, $\delta \psi$ is shown to have a complete asymptotic expansion on the asymptotically hyperbolic manifold $\cS$, as desired. An analysis of the model equations in the framework of~\cite{melrose} near $\Sigma$ was given in a similar context in the appendix of~\cite{hartog}. This completes the proof of theorem~\ref{thm3}.

\section{Summary and outlook}

In this paper, we have introduced a method for analyzing
 near null-infinity the asymptotic expansion of a
Ricci-flat, asymptotically flat metric,
in arbitrary even dimension $d \ge 4$. The method combines conformal
Gaussian null coordinates (CGNC's)  with the Ricci-flat condition. We used this
technique to obtain, starting from invariant geometric expressions
derived in~\cite{hi}, concrete expressions for the Bondi mass and
news in terms of CGNC's. These methods were then extended to
analyze the asymptotic expansion of the Witten spinor, and lead
to a proof of the positive mass theorem. The proof crucially depends on
delicate cancelations between terms in various asymptotic expansions,
and it would be interesting to have a better conceptual understanding
why these occur. Except in $d=8$, the class of ``infinities'' $\Sigma$ for which our methods work
are precisely the standard spheres $\Sigma = S^{d-2}$, while for
$d=8$, some other, more exotic, choices are also possible.

The most interesting open issue is
to generalize the positivity proof to odd dimensions, which we think should be possible. Here, CGNC's are not available, so one should perhaps use
the Bondi-type coordinates of~\cite{Tanabe:2011es, tan} instead. We also note that, by the results of~\cite{bar}, the class of possible infinities $\Sigma$ (namely, compact Riemannian $(d-2)$-manifolds
admitting a real Killing spinor) contains more  possibilities in odd $d$ apart from the
standard spheres $\Sigma=S^{d-2}$. Most of these are related to Sasakian structures~\cite{bar}. Interesting examples of
non-spherical $\Sigma$ in $d=9$ are the 7-dimensional Wallach manifolds
$\Sigma=N_{k,l}$, defined as the quotient ${\rm SU}(3)/{\rm T}$
where the torus T is embedded as $z \mapsto {\rm diag}(z^k,z^l,z^{-k-l})$.

It might also be illustrative to analyze the ``peeling theorem'' for the Weyl
tensor via CGNC's in even dimensions, complementing the analysis of~\cite{reall} done
in Bondi-type coordinates and for spherical infinities. Furthermore, it should
certainly possible to reconsider the notion of asymptotic symmetry
using~CGNC's, as was in fact done in the spherical case in~\cite{thorne}.

\vspace{2cm}

{\bf Acknowledgments:} We would like to thank A.~Ishibashi and P. Chrusciel for
discussions and perceptive comments. We would also like to thank the unknown referees 
for their useful comments. A.~T. thanks the School of Mathematics, Cardiff University, for financial support.

\appendix

\section{Einstein's equations in CGNC's}\label{appendix:A}
\label{appA}

Here we write down the vacuum Einstein's equations  in CGNC's. In
this paper, they are used most conveniently
in the form $K_g = 0$, where $K_g$ is the Schouten tensor~\eqref{schouten}
of the {\em physical} metric $g$. We hence give the coordinate expressions for
this tensor in CGNC's. $D_A$ is the derivative operator of $\gamma_{AB}$, to be distinguished
from the derivative operator $\D_A$ of $s_{AB}$, used in the body of this paper.
The inverse of $\gamma_{AB}$ is denoted as $\gamma^{AB}$. Unlike in the body of
the paper, $\gamma_{AB}$ (not $s_{AB}$) and its inverse are used
to raise and lower indices. The Ricci
tensor of $\gamma_{AB}$ is denoted by ${\mathcal R}_{AB}$.
\small
\begin{align}
\label{eq:ei-rr}
K_{rr} =
\frac{1}{d-2}\left[-\gamma^{AB}\frac{\partial^2\gamma_{AB}}{\partial r^2} + \frac{1}{2}\gamma^{CA}\gamma^{DB}\frac{\partial\gamma_{AB}}{\partial r}\frac{\partial\gamma_{CD}}{\partial r}\right]
\end{align}

\begin{align}
\label{eq:ei-ru}
&K_{ru} =
\frac{1}{(d-1)(d-2)}\left[2(d-2)(d-3)\frac{\partial(r\alpha)}{\partial r} + \frac{1}{2}(d-4)\gamma^{CA}\gamma^{DB}\frac{\partial\gamma_{AB}}{\partial r}\frac{\partial\gamma_{CD}}{\partial u} - (d-3)\gamma^{AB}\frac{\partial^2\gamma_{AB}}{\partial r\partial u}\right.\nonumber\\
& - \left.(d-3)(r\alpha)\gamma^{AB}\frac{\partial\gamma_{AB}}{\partial r} - \left(d-\frac{5}{2}\right)\gamma^{AB}\beta_A\beta_B - \gamma^{AB}\mathcal{R}_{AB} + \gamma^{AB}D_A\beta_B + \frac{1}{2}\gamma^{AB}\gamma^{CD}\frac{\partial\gamma_{CD}}{\partial u}\frac{\partial\gamma_{AB}}{\partial r}\right]\nonumber\\
& + \frac{r}{(d-1)(d-2)}\left[-2(d-2)\frac{\partial^2(r\alpha)}{\partial r^2} - (d-3)\gamma^{AB}\frac{\partial\gamma_{AB}}{\partial r}\frac{\partial(r\alpha)}{\partial r} -\frac{1}{2}(d-5)\gamma^{AB}\frac{\partial\gamma_{AB}}{\partial r}\gamma^{CD}\beta_C\beta_D\right.\nonumber\\
& + (d^2-7d+13)\gamma^{AB}\beta_A\frac{\partial\beta_B}{\partial r} - (d-2)\frac{\partial\gamma^{AB}}{\partial r}\beta_A\beta_B - (d-2)\gamma^{AB}D_A\frac{\partial\beta_B}{\partial r} + 2(r\alpha)\gamma^{AB}\frac{\partial^2\gamma_{AB}}{\partial r^2}\nonumber\\
& - \frac{3}{2}(r\alpha)\gamma^{CA}\gamma^{DB}\frac{\partial\gamma_{AB}}{\partial r}\frac{\partial\gamma_{CD}}{\partial r} - 2\gamma^{AB}\beta_B\gamma^{CD}\beta_D\frac{\partial\gamma_{AC}}{\partial r} + 2\gamma^{AB}\beta_B\gamma^{CD}D_{\left[A\right.}\frac{\partial\gamma_{|D|\left.C\right]}}{\partial r}\nonumber\\
& + \frac{1}{2}\gamma^{AB}\gamma^{CD}D_C\left(\beta_D\frac{\partial\gamma_{AB}}{\partial r}\right) + \frac{1}{2}(r\alpha)\gamma^{AB}\gamma^{CD}\frac{\partial\gamma_{CD}}{\partial r}\frac{\partial\gamma_{AB}}{\partial r} + \frac{1}{2}\gamma^{AB}\gamma^{CD}\frac{\partial\gamma_{CD}}{\partial r}D_A\beta_B\nonumber\\
& + \left.\gamma^{AB}\frac{\partial}{\partial r}\left\{D_A\beta_B\right\} - \gamma^{AB}\gamma^{CD}\left(D_D\beta_A\right)\frac{\partial\gamma_{BC}}{\partial r}\right]\nonumber\\
& + \frac{r^2}{(d-1)(d-2)}\left[-\frac{1}{2}(d-5)\gamma^{AB}\frac{\partial\gamma_{AB}}{\partial r}\gamma^{CD}\beta_C\frac{\partial\beta_D}{\partial r} - (d-2)\frac{\partial\gamma^{AB}}{\partial r}\beta_A\frac{\partial\beta_B}{\partial r} - \left(d-\frac{5}{2}\right)\gamma^{AB}\frac{\partial\beta_A}{\partial r}\frac{\partial\beta_B}{\partial r}\right.\nonumber\\
& - (d-3)\gamma^{AB}\beta_A\frac{\partial^2\beta_B}{\partial r^2} + \gamma^{AB}\frac{\partial^2\gamma_{AB}}{\partial r^2}\gamma^{CD}\beta_C\beta_D - \frac{3}{4}\gamma^{CA}\gamma^{DB}\frac{\partial\gamma_{AB}}{\partial r}\frac{\partial\gamma_{CD}}{\partial r}\gamma^{EF}\beta_E\beta_F\nonumber\\
& -\gamma^{AB}\beta_B\frac{\partial\gamma^{CD}}{\partial r}\beta_D\frac{\partial\gamma_{AC}}{\partial r} - \gamma^{AB}\beta_B\gamma^{CD}\frac{\partial\beta_D}{\partial r}\frac{\partial\gamma_{AC}}{\partial r} - \gamma^{AB}\beta_B\gamma^{CD}\beta_D\frac{\partial^2\gamma_{AC}}{\partial r}\nonumber\\
& - \frac{1}{2}\gamma^{AB}\gamma^{CD}\gamma^{EF}\beta_B\beta_F\frac{\partial\gamma_{CD}}{\partial r}\frac{\partial\gamma_{AE}}{\partial r} + \frac{1}{4}\gamma^{AB}\gamma^{CD}\frac{\partial\gamma_{CD}}{\partial r}\gamma^{EF}\beta_E\beta_F\frac{\partial\gamma_{AB}}{\partial r}  + \frac{1}{2}\gamma^{AB}\frac{\partial\gamma_{AB}}{\partial r}\frac{\partial\gamma^{CD}}{\partial r}\beta_C\beta_D\nonumber\\
& + \left.\frac{1}{2}\gamma^{AB}\gamma^{CE}\gamma^{DF}\beta_C\beta_D\frac{\partial\gamma_{AE}}{\partial r}\frac{\partial\gamma_{BF}}{\partial r} - \gamma^{AB}\gamma^{CD}\beta_D\frac{\partial\beta_A}{\partial r}\frac{\partial\gamma_{BC}}{\partial r}\right]
\end{align}

\begin{align}
\label{eq:ei-rA}
&K_{rA} =
r^{-1}\beta_A\nonumber\\
& + \frac{1}{d-2}\left[-\frac{1}{2}\beta_A\gamma^{BC}\frac{\partial\gamma_{BC}}{\partial r} + (d-4)\frac{\partial\beta_A}{\partial r} - (d-3)\gamma^{BC}\beta_C\frac{\partial\gamma_{AB}}{\partial r} - 2\gamma^{BC}D_{\left[A\right.}\frac{\partial\gamma_{|C|\left.B\right]}}{\partial r}\right]\nonumber\\
& + \frac{r}{d-2}\left[-\frac{\partial^2\beta_A}{\partial r^2} + \frac{\partial\gamma^{BC}}{\partial r}\beta_C\frac{\partial\gamma_{AB}}{\partial r} + \gamma^{BC}\frac{\partial\beta_C}{\partial r}\frac{\partial\gamma_{AB}}{\partial r} + \gamma^{BC}\beta_C\frac{\partial^2\gamma_{AB}}{\partial r^2} - \frac{1}{2}\gamma^{CD}\frac{\partial\gamma_{CD}}{\partial r}\frac{\partial\beta_A}{\partial r}\right.\nonumber\\
& + \left.\frac{1}{2}\gamma^{CD}\frac{\partial\gamma_{CD}}{\partial r}\gamma^{EF}\beta_F\frac{\partial\gamma_{AE}}{\partial r}\right]
\end{align}

\begin{align}
\label{eq:ei-uu}
&K_{uu} =
\frac{1}{d-2}\left[-\gamma^{AB}\frac{\partial^2\gamma_{AB}}{\partial u^2} + \frac{1}{2}\gamma^{CA}\gamma^{DB}\frac{\partial\gamma_{AB}}{\partial u}\frac{\partial\gamma_{CD}}{\partial u} + (r\alpha)\gamma^{AB}\frac{\partial\gamma_{AB}}{\partial u} + 2(d-2)\frac{\partial(r\alpha)}{\partial u}\right]\nonumber\\
& + \frac{r}{(d-1)(d-2)}\left[-4(d-2)(d-3)(r\alpha)\frac{\partial(r\alpha)}{\partial r} + (d-1)\frac{\partial(r\alpha)}{\partial r}\gamma^{AB}\frac{\partial\gamma_{AB}}{\partial u} - (d-1)\gamma^{AB}\frac{\partial\gamma_{AB}}{\partial r}\frac{\partial(r\alpha)}{\partial u}\right.\nonumber\\
& + 2(d-3)(r\alpha)^2\gamma^{AB}\frac{\partial\gamma_{AB}}{\partial r} + 2(d-2)(r\alpha)\gamma^{AB}D_A\beta_B - 2(d-1)(d-3)\gamma^{AB}\beta_B\frac{\partial(r\alpha)}{\partial x^A}\nonumber\\
& + 2(d-1)\gamma^{AB}D_A\left\{\frac{\partial(r\alpha)}{\partial x^B} - \frac{\partial\beta_B}{\partial u}\right\} + (10d-13)(r\alpha)\gamma^{AB}\beta_A\beta_B + 2(d-1)(d-3)\gamma^{BC}\beta_C\frac{\partial\beta_B}{\partial u}\nonumber\\
& + 3(r\alpha)\gamma^{CA}\gamma^{DB}\frac{\partial\gamma_{AB}}{\partial r}\frac{\partial\gamma_{CD}}{\partial u} - 4(r\alpha)\gamma^{AB}\frac{\partial^2\gamma_{AB}}{\partial r\partial u} + \left.2(r\alpha)\gamma^{AB}\mathcal{R}_{AB} - (r\alpha)\gamma^{AB}\gamma^{CD}\frac{\partial\gamma_{CD}}{\partial u}\frac{\partial\gamma_{AB}}{\partial r}\right]\nonumber\\
& + \frac{r^2}{(d-1)(d-2)}\left[4(d-1)(r\alpha)\frac{\partial^2(r\alpha)}{\partial r^2} - (d-1)\gamma^{AB}\frac{\partial\gamma_{AB}}{\partial r}\gamma^{CD}\beta_C\frac{\partial\beta_D}{\partial u} + (d-5)(r\alpha)\gamma^{AB}\frac{\partial\gamma_{AB}}{\partial r}\gamma^{CD}\beta_C\beta_D\right.\nonumber\\
& + 2(d-3)(r\alpha)\frac{\partial(r\alpha)}{\partial r}\gamma^{AB}\frac{\partial\gamma_{AB}}{\partial r} + (d-1)\gamma^{AB}\frac{\partial\gamma_{AB}}{\partial r}\gamma^{CD}\beta_C\frac{\partial(r\alpha)}{\partial x^D} + 2(d-1)\gamma^{AB}D_A\left(\beta_B\frac{\partial(r\alpha)}{\partial r}\right)\nonumber\\
& + 2(6d-13)(r\alpha)\gamma^{BC}\beta_C\frac{\partial\beta_B}{\partial r} + 4(d-1)\gamma^{BC}\beta_C\gamma^{AE}\beta_ED_{\left[A\right.}\beta_{\left.B\right]} - 2(d-1)\gamma^{BC}\frac{\partial\beta_C}{\partial r}\frac{\partial(r\alpha)}{\partial x^B}\nonumber\\
& - 2(d-1)(d-8)\frac{\partial(r\alpha)}{\partial r}\gamma^{AB}\beta_A\beta_B + 2(d-1)\gamma^{AB}\beta_A\beta_B\frac{\partial^2(r\alpha)}{\partial r^2} + 2(d-1)\frac{\partial\gamma^{AB}}{\partial r}\beta_A\left\{\frac{\partial(r\alpha)}{\partial x^B} - \frac{\partial\beta_B}{\partial u}\right\}\nonumber\\
& + 2(d-1)\gamma^{AB}\beta_A\left\{\frac{\partial^2(r\alpha)}{\partial r\partial x^B} - \frac{\partial^2\beta_B}{\partial u\partial r}\right\} + 2(d-2)(r\alpha)\frac{\partial\gamma^{AB}}{\partial r}\beta_A\beta_B - 4(r\alpha)^2\gamma^{AB}\frac{\partial^2\gamma_{AB}}{\partial r^2}\nonumber\\
& + 3(r\alpha)^2\gamma^{CA}\gamma^{DB}\frac{\partial\gamma_{AB}}{\partial r}\frac{\partial\gamma_{CD}}{\partial r} - 4(r\alpha)\frac{\partial^2(r\alpha)}{\partial r^2} - 2(r\alpha)\gamma^{AB}D_A\frac{\partial\beta_B}{\partial r} + 4(r\alpha)\gamma^{AB}\beta_B\gamma^{CD}\beta_D\frac{\partial\gamma_{AC}}{\partial r}\nonumber\\
& - 4(r\alpha)\gamma^{AB}\beta_B\gamma^{CD}D_{\left[A\right.}\frac{\partial\gamma_{|D|\left.C\right]}}{\partial r} - (r\alpha)\gamma^{AB}\gamma^{CD}D_C\left(\beta_D\frac{\partial\gamma_{AB}}{\partial r}\right) - (r\alpha)^2\gamma^{AB}\gamma^{CD}\frac{\partial\gamma_{CD}}{\partial r}\frac{\partial\gamma_{AB}}{\partial r}\nonumber\\
& - \left.(r\alpha)\gamma^{AB}\gamma^{CD}\frac{\partial\gamma_{CD}}{\partial r}D_A\beta_B - 2(r\alpha)\gamma^{AB}\frac{\partial}{\partial r}\left\{D_A\beta_B\right\} + 2(r\alpha)\gamma^{AB}\gamma^{CD}\left(D_D\beta_A\right)\frac{\partial\gamma_{BC}}{\partial r}\right]\nonumber\\
& + \frac{r^3}{(d-1)(d-2)}\left[(d-1)\frac{\partial(r\alpha)}{\partial r}\gamma^{AB}\frac{\partial\gamma_{AB}}{\partial r}\gamma^{CD}\beta_C\beta_D + 2(d-1)\gamma^{AB}\beta_B\frac{\partial\beta_A}{\partial r}\gamma^{CD}\beta_C\beta_D\right.\nonumber\\
& + (2d-5)(r\alpha)\gamma^{BC}\frac{\partial\beta_C}{\partial r}\frac{\partial\beta_B}{\partial r} + 4(d-1)\gamma^{BC}\frac{\partial\beta_C}{\partial r}\gamma^{AE}\beta_E D_{\left[A\right.}\beta_{\left.B\right]} + 2(d-1)\frac{\partial(r\alpha)}{\partial r}\frac{\partial\gamma^{AB}}{\partial r}\beta_A\beta_B\nonumber\\
& + 8(d-1)\frac{\partial(r\alpha)}{\partial r}\gamma^{AB}\beta_A\frac{\partial\beta_B}{\partial r} - 2(d-1)\gamma^{AB}\beta_A\frac{\partial\beta_B}{\partial r}\gamma^{CD}\beta_C\beta_D - 2(r\alpha)\gamma^{AB}\frac{\partial^2\gamma_{AB}}{\partial r^2}\gamma^{CD}\beta_C\beta_D\nonumber\\
& + \frac{3}{2}(r\alpha)\gamma^{CA}\gamma^{DB}\frac{\partial\gamma_{AB}}{\partial r}\frac{\partial\gamma_{CD}}{\partial r}\gamma^{EF}\beta_E\beta_F - 4(r\alpha)\gamma^{AB}\frac{\partial\gamma_{AB}}{\partial r}\gamma^{CD}\beta_C\frac{\partial\beta_D}{\partial r}\nonumber\\
& - 2(r\alpha)\frac{\partial\gamma^{AB}}{\partial r}\beta_A\frac{\partial\beta_B}{\partial r} - 4(r\alpha)\gamma^{AB}\beta_A\frac{\partial^2\beta_B}{\partial r^2} + 2(r\alpha)\gamma^{AB}\beta_B\frac{\partial\gamma^{CD}}{\partial r}\beta_D\frac{\partial\gamma_{AC}}{\partial r} + 2(r\alpha)\gamma^{AB}\beta_B\gamma^{CD}\frac{\partial\beta_D}{\partial r}\frac{\partial\gamma_{AC}}{\partial r}\nonumber\\
& + 2(r\alpha)\gamma^{AB}\beta_B\gamma^{CD}\beta_D\frac{\partial^2\gamma_{AC}}{\partial r} + (r\alpha)\gamma^{AB}\gamma^{CD}\gamma^{EF}\beta_B\beta_F\frac{\partial\gamma_{CD}}{\partial r}\frac{\partial\gamma_{AE}}{\partial r} - \frac{1}{2}(r\alpha)\gamma^{AB}\gamma^{CD}\frac{\partial\gamma_{CD}}{\partial r}\gamma^{EF}\beta_E\beta_F\frac{\partial\gamma_{AB}}{\partial r}\nonumber\\
& - \left.(r\alpha)\gamma^{AB}\frac{\partial\gamma_{AB}}{\partial r}\frac{\partial\gamma^{CD}}{\partial r}\beta_C\beta_D - (r\alpha)\gamma^{AB}\gamma^{CE}\gamma^{DF}\beta_C\beta_D\frac{\partial\gamma_{AE}}{\partial r}\frac{\partial\gamma_{BF}}{\partial r} + 2(r\alpha)\gamma^{AB}\gamma^{CD}\beta_D\frac{\partial\beta_A}{\partial r}\frac{\partial\gamma_{BC}}{\partial r}\right]\nonumber\\
&+ \frac{r^4}{d-2}\left[\gamma^{AB}\frac{\partial\beta_A}{\partial r}\frac{\partial\beta_B}{\partial r}\gamma^{CD}\beta_C\beta_D - \gamma^{AB}\beta_A\frac{\partial\beta_B}{\partial r}\gamma^{CD}\beta_C\frac{\partial\beta_D}{\partial r}\right]
\end{align}

\begin{align}
\label{eq:ei-uA}
&K_{uA} =
\frac{1}{d-2}\left[2(d-3)\frac{\partial(r\alpha)}{\partial x^A} + \frac{\partial\beta_A}{\partial u} + \frac{1}{2}\beta_A\gamma^{BC}\frac{\partial\gamma_{BC}}{\partial u} - 2\gamma^{BC}D_{\left[A\right.}\frac{\partial\gamma_{|C|\left.B\right]}}{\partial u} - (d-2)\gamma^{BC}\beta_C\frac{\partial\gamma_{AB}}{\partial u}\right]\nonumber\\
& + \frac{r}{(d-1)(d-2)}\left[\frac{1}{2}(d-1)\gamma^{BC}\frac{\partial\gamma_{BC}}{\partial u}\frac{\partial\beta_A}{\partial r} + (d-1)\frac{\partial^2\beta_A}{\partial u\partial r} - 2(d-1)(d-2)(r\alpha)\frac{\partial\beta_A}{\partial r} + \left(d-\frac{5}{2}\right)\beta_A\gamma^{BC}\beta_B\beta_C\right.\nonumber\\
& - 2(d-1)\frac{\partial^2(r\alpha)}{\partial r\partial x^A} + (d-1)\frac{\partial\gamma^{BC}}{\partial r}\beta_B\frac{\partial\gamma_{CA}}{\partial u} + (d-1)\gamma^{BC}\beta_B\frac{\partial^2\gamma_{CA}}{\partial r\partial u} + 2(d-1)(d-4)\gamma^{CD}\beta_C D_{\left[A\right.}\beta_{\left.D\right]}\nonumber\\
& + (d-3)(r\alpha)\gamma^{BC}\frac{\partial\gamma_{BC}}{\partial r}\beta_A - (d-1)\gamma^{BC}\frac{\partial\gamma_{BC}}{\partial r}\frac{\partial(r\alpha)}{\partial x^A} + \frac{1}{2}(d-1)\gamma^{BC}\frac{\partial\gamma_{BC}}{\partial r}\gamma^{EF}\beta_E\frac{\partial\gamma_{FA}}{\partial u}\nonumber\\
& + (d-2)\gamma^{BC}\beta_A D_B\beta_C - 2(d-1)\gamma^{CD}D_D D_{\left[A\right.}\beta_{\left.C\right]} + 2(d-1)\gamma^{BC}\frac{\partial\gamma_{CA}}{\partial r}\frac{\partial(r\alpha)}{\partial x^B} - (d-1)\gamma^{BC}\frac{\partial\gamma_{CA}}{\partial r}\frac{\partial\beta_B}{\partial u}\nonumber\\
& - \left.4\frac{\partial(r\alpha)}{\partial r}\beta_A + \frac{3}{2}\beta_A\gamma^{CE}\gamma^{DB}\frac{\partial\gamma_{EB}}{\partial r}\frac{\partial\gamma_{CD}}{\partial u} - 2\beta_A\gamma^{CD}\frac{\partial^2\gamma_{CD}}{\partial r\partial u} + \beta_A\gamma^{CD}\mathcal{R}_{CD} - \frac{1}{2}\beta_A\gamma^{BC}\gamma^{DE}\frac{\partial\gamma_{DE}}{\partial u}\frac{\partial\gamma_{BC}}{\partial r}\right]\nonumber\\
& + \frac{r^2}{(d-1)(d-2)}\left[(d-2)\beta_A\frac{\partial\gamma^{BC}}{\partial r}\beta_B\beta_C + (d-8)\beta_A\gamma^{BC}\beta_B\frac{\partial\beta_C}{\partial r} - (d-1)(d-3)\frac{\partial\beta_A}{\partial r}\gamma^{BC}\beta_B\beta_C\right.\nonumber\\
& - 2(d-1)\frac{\partial\gamma^{CD}}{\partial r}\beta_C D_{\left[A\right.}\beta_{\left.D\right]} - 2(d-1)\gamma^{CD}\frac{\partial\beta_C}{\partial r}D_{\left[A\right.}\beta_{\left.D\right]}\nonumber\\
& - 2(d-1)\gamma^{CD}\beta_C\frac{\partial}{\partial r}\left\{D_{\left[A\right.}\beta_{\left.D\right]}\right\} + \frac{1}{2}(d-5)\gamma^{BC}\frac{\partial\gamma_{BC}}{\partial r}\gamma^{EF}\beta_E\beta_F\beta_A + (d-1)(r\alpha)\gamma^{BC}\frac{\partial\gamma_{BC}}{\partial r}\frac{\partial\beta_A}{\partial r}\nonumber\\
& - (d-1)\gamma^{BC}\frac{\partial\gamma_{BC}}{\partial r}\gamma^{EF}\beta_E D_{\left[A\right.}\beta_{\left.F\right]} + (d-1)\gamma^{BC}D_B\left\{\beta_C\frac{\partial\beta_A}{\partial r}\right\}\nonumber\\
& - (d-1)\gamma^{BC}\beta_B\gamma^{EF}\beta_E\beta_F\frac{\partial\gamma_{CA}}{\partial r} + (d-1)\gamma^{BC}\beta_B\gamma^{EF}\beta_C\beta_F\frac{\partial\gamma_{EA}}{\partial r}\nonumber\\
& - 2(d-1)(r\alpha)\gamma^{BC}\frac{\partial\beta_B}{\partial r}\frac{\partial\gamma_{CA}}{\partial r} - (d-1)\gamma^{BC}\frac{\partial\beta_B}{\partial r}D_C\beta_A + 2(d-1)\gamma^{BC}\beta_B\frac{\partial\gamma_{CA}}{\partial r}\frac{\partial(r\alpha)}{\partial r}\nonumber\\
& + 2(d-1)\gamma^{BC}\frac{\partial\gamma_{CA}}{\partial r}\gamma^{EF}\beta_F D_{\left[B\right.}\beta_{\left.E\right]} - 2(r\alpha)\beta_A\gamma^{CD}\frac{\partial^2\gamma_{CD}}{\partial r^2} + \frac{3}{2}(r\alpha)\beta_A\gamma^{DB}\gamma^{EC}\frac{\partial\gamma_{BC}}{\partial r}\frac{\partial\gamma_{DE}}{\partial r}\nonumber\\
& - 2\beta_A\frac{\partial^2(r\alpha)}{\partial r^2} - 2\beta_A\gamma^{CD}\frac{\partial\gamma_{CD}}{\partial r}\frac{\partial(r\alpha)}{\partial r} - 2\beta_A\gamma^{CD}\frac{\partial\gamma_{CD}}{\partial r}\gamma^{EF}\beta_E\beta_F\nonumber\\
& - \beta_A\gamma^{BC}D_B\frac{\partial\beta_C}{\partial r} + 2\beta_A\gamma^{BC}\beta_C\gamma^{DE}\beta_E\frac{\partial\gamma_{BD}}{\partial r} - 2\beta_A\gamma^{BC}\beta_C\gamma^{DE}D_{\left[B\right.}\frac{\partial\gamma_{|E|\left.D\right]}}{\partial r}\nonumber\\
& - \frac{1}{2}\beta_A\gamma^{BC}\gamma^{DE}D_D\left(\beta_E\frac{\partial\gamma_{BC}}{\partial r}\right) - \frac{1}{2}(r\alpha)\beta_A\gamma^{BC}\gamma^{DE}\frac{\partial\gamma_{DE}}{\partial r}\frac{\partial\gamma_{BC}}{\partial r} - \frac{1}{2}\beta_A\gamma^{BC}\gamma^{DE}\frac{\partial\gamma_{DE}}{\partial r}D_B\beta_C\nonumber\\
& - \left.\beta_A\gamma^{BC}\frac{\partial}{\partial r}\left\{D_B\beta_C\right\} + \beta_A\gamma^{BC}\gamma^{DE}\left(D_E\beta_B\right)\frac{\partial\gamma_{CD}}{\partial r}\right]\nonumber\\
&+ \frac{r^3}{(d-1)(d-2)}\left[(d-1)\frac{\partial\beta_A}{\partial r}\frac{\partial\gamma^{BC}}{\partial r}\beta_B\beta_C + (d-1)\frac{\partial\beta_A}{\partial r}\gamma^{BC}\beta_B\frac{\partial\beta_C}{\partial r} + \frac{1}{2}(d-1)\gamma^{BC}\frac{\partial\gamma_{BC}}{\partial r}\gamma^{EF}\beta_E\beta_F\frac{\partial\beta_A}{\partial r}\right.\nonumber\\
& - (d-1)\gamma^{BC}\frac{\partial\beta_B}{\partial r}\gamma^{EF}\beta_E\beta_F\frac{\partial\gamma_{CA}}{\partial r} + (d-1)\gamma^{BC}\frac{\partial\beta_B}{\partial r}\gamma^{EF}\beta_C\beta_F\frac{\partial\gamma_{EA}}{\partial r} - \beta_A\gamma^{BC}\frac{\partial^2\gamma_{BC}}{\partial r^2}\gamma^{DE}\beta_D\beta_E\nonumber\\
&  + \frac{3}{4}\beta_A\gamma^{DB}\gamma^{EC}\frac{\partial\gamma_{BC}}{\partial r}\frac{\partial\gamma_{DE}}{\partial r}\gamma^{FG}\beta_F\beta_G - 2\beta_A\gamma^{BC}\frac{\partial\gamma_{BC}}{\partial r}\gamma^{DE}\beta_D\frac{\partial\beta_E}{\partial r} - \beta_A\frac{\partial\gamma^{BC}}{\partial r}\beta_B\frac{\partial\beta_C}{\partial r}\nonumber\\
& - \frac{3}{2}\beta_A\gamma^{BC}\frac{\partial\beta_B}{\partial r}\frac{\partial\beta_C}{\partial r} - 2\beta_A\gamma^{BC}\beta_B\frac{\partial^2\beta_C}{\partial r^2} + \beta_A\gamma^{BC}\beta_C\frac{\partial\gamma^{DE}}{\partial r}\beta_E\frac{\partial\gamma_{BD}}{\partial r} + \beta_A\gamma^{BC}\beta_C\gamma^{DE}\frac{\partial\beta_E}{\partial r}\frac{\partial\gamma_{BD}}{\partial r}\nonumber\\
& + \beta_A\gamma^{BC}\beta_C\gamma^{DE}\beta_E\frac{\partial^2\gamma_{BD}}{\partial r^2} + \frac{1}{2}\beta_A\gamma^{BC}\gamma^{DE}\gamma^{FG}\beta_C\beta_G\frac{\partial\gamma_{DE}}{\partial r}\frac{\partial\gamma_{BF}}{\partial r} - \frac{1}{4}\beta_A\gamma^{BC}\gamma^{DE}\frac{\partial\gamma_{DE}}{\partial r}\gamma^{FG}\beta_F\beta_G\frac{\partial\gamma_{BC}}{\partial r}\nonumber\\
& - \left.\frac{1}{2}\beta_A\gamma^{BC}\frac{\partial\gamma_{BC}}{\partial r}\frac{\partial\gamma^{DE}}{\partial r}\beta_D\beta_E - \frac{1}{2}\beta_A\gamma^{BC}\gamma^{DF}\gamma^{EG}\beta_D\beta_E\frac{\partial\gamma_{BF}}{\partial r}\frac{\partial\gamma_{CG}}{\partial r} + \beta_A\gamma^{BC}\gamma^{DE}\beta_E\frac{\partial\beta_B}{\partial r}\frac{\partial\gamma_{CD}}{\partial r}\right]
\end{align}

\begin{align}
\label{eq:ei-AB}
&K_{AB} =
r^{-1}\left[\frac{\partial\gamma_{AB}}{\partial u} - 2(r\alpha)\gamma_{AB}\right]\nonumber\\
& + \frac{1}{(d-1)(d-2)}\left[-2(d-1)\frac{\partial^2\gamma_{AB}}{\partial r\partial u} + 2(d-1)(d-3)(r\alpha)\frac{\partial\gamma_{AB}}{\partial r} + 2(d-1)\mathcal{R}_{AB} + 2(d-1)(d-3)D_{\left(A\right.}\beta_{\left.B\right)}\right.\nonumber\\
& - (d-1)\beta_A\beta_B + 2(d-1)\gamma^{CD}\frac{\partial\gamma_{D\left(A\right.}}{\partial r}\frac{\partial\gamma_{\left.B\right)C}}{\partial u} - \frac{1}{2}(d-1)\left(\gamma^{CD}\frac{\partial\gamma_{CD}}{\partial u}\frac{\partial\gamma_{AB}}{\partial r} + \gamma^{CD}\frac{\partial\gamma_{CD}}{\partial r}\frac{\partial\gamma_{AB}}{\partial u}\right)\nonumber\\
& + 4\frac{\partial(r\alpha)}{\partial r}\gamma_{AB} - \frac{3}{2}\gamma_{AB}\gamma^{EC}\gamma^{FD}\frac{\partial\gamma_{CD}}{\partial r}\frac{\partial\gamma_{EF}}{\partial u} + 2\gamma_{AB}\gamma^{CD}\frac{\partial^2\gamma_{CD}}{\partial r\partial u} + 2(r\alpha)\gamma_{AB}\gamma^{CD}\frac{\partial\gamma_{CD}}{\partial r}\nonumber\\
& - \left.\left(d^2 - 3d + \frac{1}{2}\right)\gamma_{AB}\gamma^{CD}\beta_C\beta_D - \gamma_{AB}\gamma^{CD}\mathcal{R}_{CD} + \gamma_{AB}\gamma^{CD}D_C\beta_D + \frac{1}{2}\gamma_{AB}\gamma^{CD}\gamma^{EF}\frac{\partial\gamma_{EF}}{\partial u}\frac{\partial\gamma_{CD}}{\partial r}\right] \nonumber\\
& + \frac{r}{(d-1)(d-2)}\left[-2(d-1)(r\alpha)\frac{\partial^2\gamma_{AB}}{\partial r^2} - (d-1)\gamma^{CD}D_C\left(\beta_D\frac{\partial\gamma_{AB}}{\partial r}\right) - (d-1)(r\alpha)\gamma^{CD}\frac{\partial\gamma_{CD}}{\partial r}\frac{\partial\gamma_{AB}}{\partial r}\right.\nonumber\\
& - (d-1)\gamma^{CD}\frac{\partial\gamma_{CD}}{\partial r}D_{\left(A\right.}\beta_{\left.B\right)} - 2(d-1)\frac{\partial(r\alpha)}{\partial r}\frac{\partial\gamma_{AB}}{\partial r} + (d-1)(d-4)\gamma^{CD}\beta_C\beta_D\frac{\partial\gamma_{AB}}{\partial r} - 2(d-1)\frac{\partial}{\partial r}\left\{D_{\left(A\right.}\beta_{\left.B\right)}\right\}\nonumber\\
& - 2(d-1)\beta_{\left(A\right.}\frac{\partial\beta_{\left.B\right)}}{\partial r} + 2(d-1)\gamma^{CD}\beta_D\beta_{\left(A\right.}\frac{\partial\gamma_{\left.B\right)C}}{\partial r} + 2(d-1)\gamma^{CD}\left(D_D \beta_{\left(A\right.}\right)\frac{\partial\gamma_{\left.B\right)C}}{\partial r}\nonumber\\
& + 2(d-1)(r\alpha)\gamma^{CD}\frac{\partial\gamma_{CA}}{\partial r}\frac{\partial\gamma_{DB}}{\partial r} + 2(r\alpha)\gamma_{AB}\gamma^{CD}\frac{\partial^2\gamma_{CD}}{\partial r^2} - \frac{3}{2}(r\alpha)\gamma_{AB}\gamma^{EC}\gamma^{FD}\frac{\partial\gamma_{CD}}{\partial r}\frac{\partial\gamma_{EF}}{\partial r}\nonumber\\
& + 2\frac{\partial^2(r\alpha)}{\partial r^2}\gamma_{AB} + 2\gamma_{AB}\gamma^{CD}\frac{\partial\gamma_{CD}}{\partial r}\frac{\partial(r\alpha)}{\partial r} + 2\gamma_{AB}\gamma^{CD}\frac{\partial\gamma_{CD}}{\partial r}\gamma^{EF}\beta_E\beta_F + 7\gamma_{AB}\gamma^{CD}\beta_C\frac{\partial\beta_D}{\partial r} + \gamma_{AB}\frac{\partial\gamma^{CD}}{\partial r}\beta_C\beta_D\nonumber\\
& + \gamma_{AB}\gamma^{CD}D_C\frac{\partial\beta_D}{\partial r} - 2\gamma_{AB}\gamma^{CD}\beta_D\gamma^{EF}\beta_F\frac{\partial\gamma_{CE}}{\partial r} + 2\gamma_{AB}\gamma^{CD}\beta_D\gamma^{EF}D_{\left[C\right.}\frac{\partial\gamma_{|F|\left.E\right]}}{\partial r}\nonumber\\
& + \frac{1}{2}\gamma_{AB}\gamma^{CD}\gamma^{EF}D_E\left(\beta_F\frac{\partial\gamma_{CD}}{\partial r}\right) + \frac{1}{2}(r\alpha)\gamma_{AB}\gamma^{CD}\gamma^{EF}\frac{\partial\gamma_{EF}}{\partial r}\frac{\partial\gamma_{CD}}{\partial r} + \frac{1}{2}\gamma_{AB}\gamma^{CF}\gamma^{EF}\frac{\partial\gamma_{EF}}{\partial r}D_C\beta_D\nonumber\\
& + \left.\gamma_{AB}\gamma^{CD}\frac{\partial}{\partial r}\left\{D_C\beta_D\right\} - \gamma_{AB}\gamma^{CD}\gamma^{EF}\left(D_F\beta_C\right)\frac{\partial\gamma_{DE}}{\partial r}\right]\nonumber\\
& + \frac{r^2}{(d-1)(d-2)}\left[-\frac{1}{2}(d-1)\gamma^{CD}\frac{\partial\gamma_{CD}}{\partial r}\gamma^{EF}\beta_E\beta_F\frac{\partial\gamma_{AB}}{\partial r} - (d-1)\frac{\partial\gamma_{AB}}{\partial r}\frac{\partial\gamma^{CD}}{\partial r}\beta_C\beta_D\right.\nonumber\\
& - 2(d-1)\frac{\partial\gamma_{AB}}{\partial r}\gamma^{CD}\beta_C\frac{\partial\beta_D}{\partial r} - (d-1)\frac{\partial^2\gamma_{AB}}{\partial r^2}\gamma^{CD}\beta_C\beta_D - (d-1)\frac{\partial\beta_A}{\partial r}\frac{\partial\beta_B}{\partial r}\nonumber\\
& - (d-1)\gamma^{CE}\gamma^{DF}\beta_C\beta_D\frac{\partial\gamma_{AE}}{\partial r}\frac{\partial\gamma_{BF}}{\partial r} + 2(d-1)\gamma^{CD}\beta_D\frac{\partial\beta_{\left(A\right.}}{\partial r}\frac{\partial\gamma_{\left.B\right)C}}{\partial r} + (d-1)\gamma^{CD}\gamma^{EF}\beta_E\beta_F\frac{\partial\gamma_{CA}}{\partial r}\frac{\partial\gamma_{DB}}{\partial r}\nonumber\\
& + \gamma_{AB}\gamma^{CD}\frac{\partial^2\gamma_{CD}}{\partial r^2}\gamma^{EF}\beta_E\beta_F - \frac{3}{4}\gamma_{AB}\gamma^{EC}\gamma^{FD}\frac{\partial\gamma_{CD}}{\partial r}\frac{\partial\gamma_{EF}}{\partial r}\gamma^{GH}\beta_G\beta_H + 2\gamma_{AB}\gamma^{CD}\frac{\partial\gamma_{CD}}{\partial r}\gamma^{EF}\beta_E\frac{\partial\beta_F}{\partial r}\nonumber\\
& + \gamma_{AB}\frac{\partial\gamma^{CD}}{\partial r}\beta_C\frac{\partial\beta_D}{\partial r} + \frac{3}{2}\gamma_{AB}\gamma^{CD}\frac{\partial\beta_C}{\partial r}\frac{\partial\beta_D}{\partial r} + 2\gamma_{AB}\gamma^{CD}\beta_C\frac{\partial^2\beta_D}{\partial r^2} - \gamma_{AB}\gamma^{CD}\beta_D\frac{\partial\gamma^{EF}}{\partial r}\beta_F\frac{\partial\gamma_{CE}}{\partial r}\nonumber\\
& - \gamma_{AB}\gamma^{CD}\beta_D\gamma^{EF}\frac{\partial\beta_F}{\partial r}\frac{\partial\gamma_{CE}}{\partial r} - \gamma_{AB}\gamma^{CD}\beta_D\gamma^{EF}\beta_F\frac{\partial^2\gamma_{CE}}{\partial r^2} - \frac{1}{2}\gamma_{AB}\gamma^{CD}\gamma^{EF}\gamma^{GH}\beta_D\beta_H\frac{\partial\gamma_{EF}}{\partial r}\frac{\partial\gamma_{CG}}{\partial r}\nonumber\\
& + \frac{1}{4}\gamma_{AB}\gamma^{CD}\gamma^{EF}\frac{\partial\gamma_{EF}}{\partial r}\gamma^{GH}\beta_G\beta_H\frac{\partial\gamma_{CD}}{\partial r} + \frac{1}{2}\gamma_{AB}\gamma^{CD}\frac{\partial\gamma_{CD}}{\partial r}\frac{\partial\gamma^{EF}}{\partial r}\beta_E\beta_F\nonumber\\
& + \left.\frac{1}{2}\gamma_{AB}\gamma^{CD}\gamma^{EG}\gamma^{FH}\beta_E\beta_F\frac{\partial\gamma_{CG}}{\partial r}\frac{\partial\gamma_{DH}}{\partial r} - \gamma_{AB}\gamma^{CD}\gamma^{EF}\beta_F\frac{\partial\beta_C}{\partial r}\frac{\partial\gamma_{DE}}{\partial r}\right]
\end{align}


\end{document}